\documentclass[11pt]{cernrep}
\usepackage{epsfig,axodraw}
\usepackage{graphicx}
\begin{document}

\newcommand{\DEF}[1]{{#1}\index{#1}}
\newcommand{\D}[1]{\mbox{$ \Delta\! {\rm #1}$~}}
\newcommand{\KL}{\mbox{$\kl$~}}
\newcommand{\KS}{\mbox{$\ks$~}}
\newcommand{\bKS}{$\mathrm \mathbf K_{\mathrm \mathbf S}$~}
\newcommand{\N}[1]{\mbox{${\cal N}_{#1}$}}
\newcommand{\R}{\mbox{$\cal R$~}}
\newcommand{\TAMcons}[3]{
            {#1}({\rm TAM}_{\rm n}#2\; \& \;{\rm TAM}_{\rm n+1}#3)} 
\newcommand{\aGT}{\gtrsim}
\newcommand{\aLT}{\lesssim}
\newcommand{\alias}[1]{~\mbox{({\em #1})}\index{#1}~}
\newcommand{\bbb}{\mbox{$\rm \bz\bzb$~}}
\newcommand{\boh}{}
\newcommand{\bzb}{\mbox{$\rm \bar{B}^{0}$}~}
\newcommand{\bz}{\mbox{$\rm B^{0}$}~}
\newcommand{\cfr}[1]{~(\S~\ref{sec:#1})}
\newcommand{\chiq}{\chi^{2}}
\newcommand{\comm}[1]{}
\newcommand{\colla}{\vspace*{10cm}}
\newcommand{\ddb}{\mbox{$\rm \dz\dzb$~}}
\newcommand{\dr}[1]{\mbox{${\cal R}_{#1}$}}
\newcommand{\dtarg}{\mbox{$d_{\rm targ}$}~}
\newcommand{\dzb}{\mbox{$\rm \bar{D}^{0}$}~}
\newcommand{\dz}{\mbox{$\rm D^{0}$}~}
\newcommand{\ek}{\mbox{$E_{ K}$}~}
\newcommand{\eoe}{\mbox{${\rm Re}(\epsi'/\epsi)$}~}
\newcommand{\epsi}{\varepsilon}
\newcommand{\eqz}[1]{~(Eq.~\ref{eq:#1})}
\newcommand{\fez}{f_{\rm TAM=0}}
\newcommand{\fig}[1]{~(Fig.~\ref{fig:#1})}
\newcommand{\ket}{\mbox{$\mathrm K_{e3}$}~}
\newcommand{\kkb}{\mbox{$\rm \kz\kzb$~}}
\newcommand{\kl}{{ K_{\rm L}}}
\newcommand{\kmut}{\mbox{$\mathrm K_{\mu3}$}~}
\newcommand{\kpit}{\mbox{$\mathrm K_{\pi3}$}~}
\newcommand{\ks}{{K_{\rm S}}}
\newcommand{\kto}[1]{\mbox{$\rm \kz\rightarrow{#1}$~}}
\newcommand{\kzb}{\mbox{$ \bar{K}^{0}$}~}
\newcommand{\kz}{\mbox{$  K^{0}$}~}
\newcommand{\mini}{\mbox{\rm MINI-DST}~}
\newcommand{\pauli}[1]{\mbox{\large \boldmath $\sigma_{#1}$}}
\newcommand{\pmi}{{\pi^{-}}}
\newcommand{\ppl}{{\pi^{+}}}
\newcommand{\pz}{{\pi^{0}}}
\newcommand{\rell}{\mbox{$\mathrm R_{\mathrm ell}$}~}
\newcommand{\res}[1]{~(\ref{res:#1})}
\newcommand{\state}[1]{\left|{#1}\right>}
\newcommand{\tab}[1]{~(Tab.~\ref{tab:#1})}
\newcommand{\teq}[1]{{\rm TAM\!\!=\!\!#1}}
\newcommand{\tez}{$\teq{0}$~}
\newcommand{\tgt}[1]{{\rm TAM\!\!>\!\!#1}}
\newcommand{\tgz}{$\tgt{0}$~}
\newcommand{\tz}{\mbox{$t_0$}~}
\newcommand{\zv}{\mbox{$z_v$}~}
%
\newcommand{\piee} {\pi^0 e^+ e^-}
\newcommand{\ppee} {\pi^+ \pi^- e^+ e^-}
\newcommand{\pimumu} {\pi ^0 \mu ^+ \mu ^-}
\newcommand{\gee} {\gamma e^+ e^-}
\newcommand{\gmumu} {\mathrm{\gamma \mu ^+ \mu ^-}}
\newcommand{\gaga} {\mathrm{\gamma \gamma}}
\newcommand{\pigg} {\mathrm{\pi ^0 \gamma \gamma}}
\newcommand{\threepio} {\mathrm{\pi ^0 \pi ^0 \pi ^0}}
\newcommand{\pipigg} {\mathrm{\pi ^0 \pi ^0 \gamma \gamma}}
\newcommand{\pipig} {\mathrm{\pi ^0 \pi ^0 \gamma}}
\newcommand{\ppg} {\mathrm{\pi ^+ \pi ^- \gamma}}
\newcommand{\threepi} {\mathrm{\pi ^+ \pi ^- \pi ^0}}
\newcommand{\mumu} {\mathrm{\mu ^+ \mu ^-}}
\newcommand{\ee} {e^+ e^-}
\newcommand{\KLGG} {\kl \rightarrow \gaga}
\newcommand{\KSLGG} {K_{S,L} \rightarrow \gaga}
\newcommand{\KLPGG} {\kl \rightarrow \pigg}
\newcommand{\KLPPG} {\kl \rightarrow \pipig}
\newcommand{\KLPPGG} {\kl \rightarrow \pipigg}
\newcommand{\KSLPGG} {K_{S,L} \rightarrow \pigg}
\newcommand{\KSLPPGG} {K_{S,L} \rightarrow \pipigg}
\newcommand{\KSEEGG} {\ks \rightarrow \ee \gaga}
\newcommand{\KSEEMM} {\ks \rightarrow \ee \mumu}
\newcommand{\KLEEMM} {\kl \rightarrow \ee \mumu}
\newcommand{\KSEEEE} {\ks \rightarrow \ee \ee}
\newcommand{\KLEEEE} {\kl \rightarrow \ee \ee}
\newcommand{\KLPEE} {\kl \rightarrow \piee}
\newcommand{\KLPPEE} {\kl \rightarrow \ppee}
\newcommand{\KLEEGG} {\kl \rightarrow \ee \gaga}
\newcommand{\KLMMGG} {\kl \rightarrow \mumu \gaga}
\newcommand{\KLPMM} {\kl \rightarrow \pimumu}
\newcommand{\KLPP} {\kl \rightarrow \pi^0 \pi^0}
\newcommand{\KLPPP} {\kl \rightarrow \threepio}
\newcommand{\KLPPPC} {\kl \rightarrow \threepi}
\newcommand{\KSGG} {\ks \rightarrow \gaga}
\newcommand{\KSPPGG} {\ks \rightarrow \pipigg}
\newcommand{\KSPGG} {\ks \rightarrow \pigg}
\newcommand{\KSPPG} {\ks \rightarrow \pipig}
\newcommand{\KSPEE} {\ks \rightarrow \piee}
\newcommand{\KSPPEE} {\ks \rightarrow \ppee}
\newcommand{\KSPMM} {\ks \rightarrow \pimumu}
\newcommand{\KSPP} {\ks \rightarrow \pi \pi}
\newcommand{\KSPZPZ} {\ks \rightarrow \pi^0 \pi^0}
\newcommand{\KSPPP} {\ks \rightarrow \threepio}
\newcommand{\CP} {\mathrm{\hspace{-1.5mm} \diagup \hspace{-4.5mm} C\!P}}
\newcommand{\BR} {\mathrm{BR}}
\newcommand{\Imm} {\frak{Im}}
\newcommand{\Ree} {\frak{Re}}
\newcommand{\kbar} {\mathrm{\overline{K}^{_{\scriptstyle 0}}}}
%
\newcommand{\PreserveBackslash}[1]{\let\temp=\\#1\let\\=\temp}
\newcommand{\E}[1] {\mathrm{\times 10^{#1}}}

\newcommand{\definmath}[2] {\def#1{\ifmmode#2\else\ensuremath{\mathrm{#2}}
\fi}}

\newcommand{\epe} {\mathrm{\epsilon'/\epsilon}}
\newcommand{\delp} {\mathrm{\Delta p}}
\newcommand{\ddpk} {\mathrm{\Delta p_K}}


\pagestyle{plain}
\begin{titlepage}
\nopagebreak
{\flushright{
        CERN-TH/2001-175\\
         hep-ph/0107046\\
          July 2001\\
}        
}
\vfill
\begin{center}
{\LARGE { \bf \sc KAON PHYSICS\\ WITH A HIGH-INTENSITY PROTON DRIVER}

}
\vfill
  A.~Belyaev~$^{a}$, 
  G.~Buchalla~$^{b}$ (convener), 
  A.~Ceccucci~$^{b}$, 
  M.~Chizhov~$^{b,c}$, 
  G.~D'Ambrosio~$^{d}$, 
  A.~Dorokhov~$^{e}$, 
  J.~Ellis~$^{b}$, 
  M.~E. G\'omez~$^{f}$, 
  T.~Hurth~$^{b}$, 
  G.~Isidori~$^{b}$, 
  G.~Kalmus~$^{g}$,
  S.~Lola~$^{b}$, 
  K.~Zuber~$^{h}$
 \\

\vspace*{1cm}

{\small
  $^{a}$~Skobeltsyn Institute for Nuclear Physics, MSU, Moscow, Russia\\
  $^{b}$~CERN, Geneva, Switzerland\\
  $^{c}$~Centre for Space Research and Technologies, Faculty of
         Physics, University of Sofia, Sofia, Bulgaria\\
  $^{d}$~INFN, Sezione di Napoli, Naples, Italy\\
  $^{e}$~Bogoliubov Laboratory for Theoretical Physics, JINR, Dubna, Russia\\
  $^{f}$~CFIF, Departamento de Fisica, Instituto Superior T\'ecnico,
         Lisboa, Portugal\\
  $^{g}$~Cavendish Laboratory, University of Cambridge, 
  Cambridge, U.K.\\
  $^{h}$~Lehrstuhl f\"ur Experimentelle Physik IV, Universit\"at Dortmund,
  Dortmund, Germany}

\end{center}                                   
\nopagebreak
\vskip 1cm
\begin{abstract} 
We study opportunities for future high-precision experiments in kaon
physics using a high-intensity proton driver, which could be
part of the front-end of a muon storage ring complex.
We discuss in particular the rare decays $K_L\to\pi^0\nu\bar\nu$,
$K^+\to\pi^+\nu\bar\nu$, $K_L\to\pi^0e^+e^-$, and lepton-flavour
violating modes such as $K_L\to\mu e$ and $K\to\pi\mu e$.
The outstanding physics potential and long-term interest of these
modes is emphasized. We review status and prospects of current and 
planned experiments for the processes under consideration,
and indicate possible improvements and strategies towards
achieving the necessary higher sensitivity.
Finally, we outline the machine requirements needed to perform
these high-precision kaon experiments in the context
of a muon storage ring facility. 
\end{abstract}                                                
\vskip 1cm

\begin{center}
Kaon Physics Working Group Report\\ ECFA Studies on Neutrino
Factory and Muon Storage Rings at CERN  
\end{center}

\vskip 0.6cm
CERN-TH/2001-175\hfill   
\vfill       
\end{titlepage}

\tableofcontents

\title{KAON PHYSICS WITH A HIGH-INTENSITY PROTON DRIVER}

\author{A.~Belyaev~$^{a}$, 
  G.~Buchalla~$^{b}$ (convener), 
  A.~Ceccucci~$^{b}$, 
  M.~Chizhov~$^{b,c}$, 
  G.~D'Ambrosio~$^{d}$, 
  A.~Dorokhov~$^{e}$, 
  J.~Ellis~$^{b}$, 
  M.~E. G\'omez~$^{f}$, 
  T.~Hurth~$^{b}$, 
  G.~Isidori~$^{b}$, 
  G.~Kalmus~$^{g}$,
  S.~Lola~$^{b}$, 
  K.~Zuber~$^{h}$}

\institute{
  $^{a}$~Skobeltsyn Institute for Nuclear Physics, MSU, Moscow, Russia\\
  $^{b}$~CERN, Geneva, Switzerland\\
  $^{c}$~Centre for Space Research and Technologies, Faculty of
         Physics, University of Sofia, Sofia, Bulgaria\\
  $^{d}$~INFN, Sezione di Napoli, Naples, Italy\\
  $^{e}$~Bogoliubov Laboratory for Theoretical Physics, JINR, Dubna, Russia\\
  $^{f}$~CFIF, Departamento de Fisica, Instituto Superior T\'ecnico,
        Lisboa, Portugal\\
  $^{g}$~Cavendish Laboratory, University of Cambridge, 
        Cambridge, U.K.\\
  $^{h}$~Lehrstuhl f\"ur Experimentelle Physik IV, Universit\"at Dortmund,
        Dortmund, Germany}

\maketitle

\begin{abstract}
We study opportunities for future high-precision experiments in kaon
physics using a high-intensity proton driver, which could be
part of the front-end of a muon storage ring complex.
We discuss in particular the rare decays $K_L\to\pi^0\nu\bar\nu$,
$K^+\to\pi^+\nu\bar\nu$, $K_L\to\pi^0e^+e^-$, and lepton-flavour
violating modes such as $K_L\to\mu e$ and $K\to\pi\mu e$.
The outstanding physics potential and long-term interest of these
modes is emphasized. We review status and prospects of current and 
planned experiments for the processes under consideration,
and indicate possible improvements and strategies towards
achieving the necessary higher sensitivity.
Finally, we outline the machine requirements needed to perform
these high-precision kaon experiments in the context
of a muon storage ring facility.
\end{abstract}

\section{INTRODUCTION}

{\it G. Buchalla}

\subsection{Preliminary remarks}

There are essentially two frontiers in particle physics:
one striving for higher energies to access new degrees of freedom
directly, the other aiming for higher precision in the
study of rare processes. 
Each addresses the problem of achieving         
a better understanding of fundamental interactions from a different
perspective and in a complementary way, and both are indispensable
to obtain the complete picture.

The indirect method of high-precision measurements at
low energies is exemplified by the detailed study of kaon decays,
which has proved to be of the utmost importance for the development
of particle physics.
Among the landmark results in this field have been the concept of
strangeness,
leading to the quark model, which in turn provided the basis
for QCD; the first hint of parity violation, pointing the way to the
chiral nature of the weak gauge forces; the violation of CP, defining
an absolute matter-antimatter asymmetry and a subtle connection
to the three-generation structure of matter; and, last but not least,
the characteristic suppression pattern for flavour-changing
neutral currents (FCNC) in $K_L\to\mu^+\mu^-$ or $K^0$--$\bar K^0$
mixing, which suggested the charm quark and the GIM structure
of flavour dynamics.

These examples illustrate how rare processes with kaons 
can provide indirect and yet crucial information
on fundamental physics that is difficult to access in any other way.
The fact that kaons can be copiously produced and that they have
rather long lifetimes, with a large hierarchy between $K_L$ and
$K_S$, are key elements in this respect.

Beyond the achievements of the past kaon physics offers clear
perspectives for future progress. Prominent opportunities are the
rare decays $K\to\pi\nu\bar\nu$, $K_L\to\pi^0e^+e^-$ or
$K_L\to\mu e$. These processes allow us to perform high-precision
tests of Standard Model (SM) flavour physics, including the CKM
mechanism for CP violation, and define very sensitive probes of
new physics.
The pursuit of these investigations is important because
the physics of flavour represents the least understood and least
tested sector of the SM. Kaon probes are necessary
to complement both direct searches for new physics as well as 
the results on flavour dynamics from $b$ physics and similar fields.

The purpose of this document is to review particular
possibilities in kaon physics with a long-term perspective.
These require, on the one hand, a longer-term experimental effort,
but will yield, on the other, results of the highest interest, which
are competitive even on this longer time scale. Their interest may
be an important element in the choice of the energy of the proton
driver for muon storage rings. For this reason, in the first part of
this document
we highlight and motivate the cases of particular interest
for the very high intensity kaon beams that could be obtained from such a 
proton driver. These have been the focus at this
workshop, and we shall elaborate briefly on the studies performed
with this perspective.

In the second part of this article, we discuss the experimental
requirements for this type of precision kaon physics, including strategies
for the detection of the rare modes under consideration, a comparison with
related efforts at other laboratories, and accelerator aspects.  We
finally present a summary and our conclusions.

\subsection{Overview}

In Table 1 we list a selection of rare kaon processes
that are of special interest for future studies of flavour physics.
The chosen examples are intended to represent the typical
categories one might distinguish for rare kaon decays with sensitivity
to short-distance flavour dynamics: tests of the SM FCNC, the
CKM mechanism and CP violation; searches for (charged-)lepton-flavour
violation; probes of discrete symmetries by measuring the
polarization of muons from $K$ decays. All three classes might
reveal the presence of new physics, the first class by yielding
results for SM parameters in conflict with other determinations,
the latter two by merely giving a non-zero signal above the
(exceedingly small) level expected in the SM.
More examples have been discussed in the literature. For
further information we refer to the review articles 
\cite{LV,RW,WW,BR,BK,DI}.
Here we would like to emphasize some decays where the
theoretical motivation is particularly strong, which have been
the main focus of the present study:

\begin{itemize}
\item
$K_L\to\pi^0\nu\bar\nu$, (and also $K^+\to\pi^+\nu\bar\nu$)
\item
$K_L\to\pi^0e^+e^-$
\item
$K_L\to\mu e$
\end{itemize}
The significance of these modes is such that they are sufficient
to motivate a dedicated kaon physics program. Further opportunities
may then also be pursued later on, once a kaon facility is in place.

\begin{center}
{\small Table 1: Overview of rare kaon decays.
(The SM expectation quoted for $B(K_L\to\pi^0e^+e^-)$ refers only to the
contribution from direct CP violation, while the experimental
limit refers to the total branching fraction.)}
\vskip0.2cm
\begin{tabular}{|c|c|c|c|}
\hline
Observable & Physics & SM &  Experiment \\
\hline
\hline
$B(K_L\to\pi^0\nu\bar\nu)$ & CPV, $\eta$, $J_{CP}$ &
   $(2.8\pm 1.1)\times 10^{-11}$ \cite{AJB99} & 
   $< 5.9\times 10^{-7}$ \cite{KTeV_p0nunu}\\
\hline
$B(K^+\to\pi^+\nu\bar\nu)$ & $|V_{td}|$ &
  $(0.8\pm 0.3)\times 10^{-10}$ \cite{AJB99} & 
  $1.5^{+3.4}_{-1.2}\times 10^{-10}$ \cite{ADL}\\

\hline
$B(K_L\to\pi^+\pi^-\nu\bar\nu)$ & CKM &
   $(2.8\pm 0.8)\times 10^{-13}$ \cite{LVG} &   -- \\
\hline
$B(K^+\to\pi^+\pi^0\nu\bar\nu)$ & CKM &
   $(1-2)\times 10^{-14}$ \cite{LVG} & 
   $< 4.3\times 10^{-5}$ \cite{ADL:kppnn}\\

\hline
$B(K_L\to\pi^0e^+e^-)$ & CPV, $\eta$, $J_{CP}$ &
   $(4.6\pm 1.8)\times 10^{-12}$ \cite{AJB99} & 
   $< 5.1\times 10^{-10}$ \cite{KTeV_p0ee}\\

\hline
\hline
$B(K_L\to\mu e)$ & LFV, new physics & -- & 
   $< 4.7\times 10^{-12}$ \cite{BNL871}\\

\hline
$B(K^+\to\pi^+\mu^+ e^-)$ & LFV, new physics & -- & 
   $< 2.8\times 10^{-11}$ \cite{BNL865}\\

\hline
$B(K_L\to\pi^0\mu e)$ & LFV, new physics & -- & 
   $< 4.4\times 10^{-10}$ \cite{KTeV:lfv}\\

\hline
$B(K^+\to\pi^-\mu^+\mu^+)$ & $\Delta L_\mu=2$, new physics & -- & 
   $< 3.0\times 10^{-9}$ \cite{BNL865b}\\

\hline
\hline

$P^\mu_\perp(K^+\to\pi^0\mu^+\nu)$ & new CPV scalar int. & $\sim 10^{-6}$ & 
   $< 5\times 10^{-3}$ \cite{KEK246}\\

\hline
\end{tabular}
\end{center}

\newpage

\section{THE GOLDEN MODES: $K_L\to\pi^0\nu\bar\nu$ AND 
$K^+\to\pi^+\nu\bar\nu$}
                   
{\it G. Buchalla, T. Hurth}

\begin{figure}[t]
\hspace*{2.5cm}\epsfig{figure=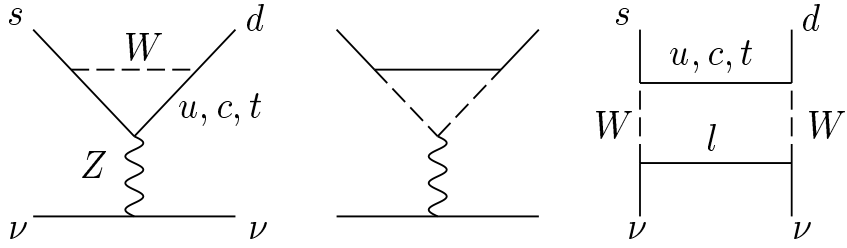,width=11.5cm,height=3.5cm}
\caption{Leading-order electroweak diagrams contributing to
           $K\to\pi\nu\bar\nu$ in the SM.}
\label{fig:kpnn}
\end{figure}

\noindent     
The decay modes $K\to\pi\nu\bar\nu$ are flavour-changing neutral
current transitions, which are induced in the SM at
one-loop order through the diagrams shown in Fig.~\ref{fig:kpnn}. 
                                       
The rare decay $K_L\to\pi^0\nu\bar\nu$ is one of the most attractive
processes to study the physics of flavour.
In particular, $K_L\to\pi^0\nu\bar\nu$ is a
manifestation of large direct CP violation in the SM.
A small effect from indirect CP violation
related to the kaon $\varepsilon$ parameter contributes
less than $\sim 1\%$ to the branching ratio, and is therefore negligible.
In addition ,
$K_L\to\pi^0\nu\bar\nu$ can be calculated as a function of fundamental
SM parameters with exceptionally small theoretical error.
The main reasons are the hard GIM suppression of long-distance
contributions \cite{RS,BI}, and the semileptonic character, which allows
us to extract the hadronic matrix element
$\langle\pi^0|(\bar sd)_V|K^0\rangle$ from $K^+\to\pi^0e\nu$ decay
using isospin symmetry. As a consequence, the $K_L\to\pi^0\nu\bar\nu$ 
amplitude is
based on a purely short-distance-dominated flavour-changing neutral
current matrix element, which is reliably calculable in perturbation
theory. 
The
CP properties help to improve further the theoretical accuracy, rendering 
even the charm contribution completely negligible, so that the clean
top contribution fully dominates the decay.
Next-to-leading QCD effects have been calculated and reduce
the leading-order scale ambiguity of $\sim \pm 10\%$ to an
essentially negligible $\sim \pm1\%$ \cite{BB2}.
Isospin-breaking corrections in the extraction of the matrix element
have also been evaluated, and lead to an overall reduction of the
branching ratio by $5.6\%$ \cite{MP}. Uncertainties from
higher-order electroweak corrections are likewise at the level of
a percent \cite{BB97}. In total, the theoretical uncertainty
in $K_L\to\pi^0\nu\bar\nu$ is below $3\%$.
Consequently, on the order of a 1000 background-free
events could still be used without being limited by theoretical 
uncertainties. The theoretically very clean relationship between
the $K_L\to\pi^0\nu\bar\nu$ branching fraction and fundamental
SM parameters reads
\begin{equation}\label{bklpnn}
B(K_L\to\pi^0\nu\bar\nu)=
1.80\times 10^{-10}\left(\frac{{\rm Im}\lambda_t}{\lambda^5} X(x_t)\right)^2
=4.16\times 10^{-10}\, 
\eta^2 A^4\left(\frac{\bar m_t(m_t)}{167 {\rm GeV}}\right)^{2.30}
\end{equation}
Here $\lambda_t\equiv V^*_{ts} V_{td}$; $\lambda=0.22$, 
$A\equiv V_{cb}/\lambda^2$ and $\eta$
are Wolfenstein parameters of the CKM matrix, 
and $X(x_t)$ is a function of the
top-quark ${\overline{MS}}$-mass $\bar m_t(m_t)$:
$x_t\equiv (\bar m_t(m_t)/M_W)^2$.

The CP-conserving mode $K^+\to\pi^+\nu\bar\nu$ is also of great
interest, being sensitive to $|V_{td}|$. Compared to the neutral
channel, $K^+\to\pi^+\nu\bar\nu$ has a
slightly larger theoretical uncertainty, which is due to the charm 
contribution that is non-negligible in this case. Explicit
expressions can be found in the third reference of \cite{BB2}. 

The study of $K\to\pi\nu\bar\nu$ can give crucial information
for testing the CKM picture of flavour mixing. This information is
complementary to the results expected from $B$ physics and is much
needed to provide the overdetermination of the unitarity triangle
necessary for a decisive test. 
Let us briefly illustrate  some specific opportunities.

The quantity $B(K_L\to\pi^0\nu\bar\nu)$ offers probably the
best accuracy in determining 
$|\mbox{Im} V^*_{ts}V_{td}|$ or,
equivalently, the Jarlskog parameter
$J_{CP}=\mbox{Im}(V^*_{ts}V_{td}V_{us}V^*_{ud})$, the invariant measure
of CP violation in the SM.
The prospects here are even better than for $B$ physics at the LHC
\cite{BB96}.
For example, a $10\%$ measurement
$B(K_L\to\pi^0\nu\bar\nu)=(3.0\pm 0.3)\times 10^{-11}$ would directly
give ${\rm Im}\lambda_t=(1.38\pm 0.07)\times 10^{-4}$, a remarkably 
precise result.
The SM expectation for the branching ratio \cite{AJB99} is 
$(2.8\pm 1.1)\times 10^{-11}$, where the uncertainty is due to our
imprecise knowledge of CKM parameters. 
The current upper bound
from direct searches \cite{KTeV_p0nunu} is $5.9\times 10^{-7}$. An indirect
upper bound, using the current limit on $B(K^+\to\pi^+\nu\bar\nu)$
\cite{ADL} and isospin symmetry, can be placed \cite{GN} at 
$2\times 10^{-9}$.

\begin{figure}[h]
\hspace*{4.5cm}\epsfig{figure=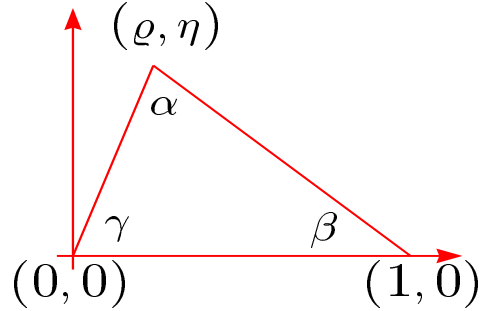,width=6.75cm,height=4.5cm}
\caption{The Wolfenstein parameters $\varrho$ and $\eta$ of the CKM
matrix, and the unitarity triangle.}
\label{fig:utre}
\end{figure}

A measurement of $B(K^+\to\pi^+\nu\bar\nu)$ to $10\%$ accuracy
can be expected to determine $|V_{td}|$ with about the same precision.
Combining $10\%$ measurements of both $K_L\to\pi^0\nu\bar\nu$
and $K^+\to\pi^+\nu\bar\nu$ determines the unitarity
triangle parameter $\sin 2\beta$, as seen in Fig.~\ref{fig:utre},
with an uncertainty of about $\pm 0.07$, comparable to the 
precision obtainable for
the same quantity from CP violation in $B\to J/\Psi K_S$
before the LHC era \cite{BB4}:
\begin{equation}\label{sin2bk}
\sin 2\beta=\frac{2 r}{1+r^2}\quad
r=\sqrt{\sigma}\frac{\sqrt{\sigma(B_+-B_L)}-P_0(K^+)}{\sqrt{B_L}}
\end{equation} 
where $\sigma=(1-\lambda^2/2)^{-2}$,
$P_0(K^+)=0.42\pm 0.06$ is the internal charm contribution
to $K^+\to\pi^+\nu\bar\nu$, which is known to next-to-leading
order in QCD, and $B_+$ and $B_L$ represent the reduced branching
ratios
$B_+=B(K^+\to\pi^+\nu\bar\nu)/(4.11\times 10^{-11})$ and
$B_L=B(K_L\to\pi^0\nu\bar\nu)/(1.80\times 10^{-10})$.

\noindent
The time-integrated CP violating asymmetry in $B_d\to\Psi K_S$
is given by 
\begin{equation}
{\cal A}_{CP}(\Psi K_S)=-\sin(2\beta)\ x_d/(1+x^2_d),
\label{ACP}
\end{equation}
where $x_d=\Delta M_{B_d}/\Gamma_{B_d}$ gives the size of
$B_d -\bar B_d$ mixing. 
Both determinations of $\sin 2\beta$, the one from (\ref{sin2bk})
and the one from $B_d\to\Psi K_S$, have to coincide if the
SM is valid. This implies the relation 
\begin{equation}\label{bkrel}
\frac{2 r(B_+, B_L)}{1+r^2(B_+, B_L)}=
-{\cal A}_{CP}(\Psi K_S)\frac{1+x^2_d}{x_d}
\end{equation}
which represents a crucial test of the SM.
As was stressed in \cite{BB4}, all quantities in the
`golden relation' (\ref{bkrel}) -- except for the (calculable)
$P_0(K^+)$ -- can be directly measured by experiment,
and the relation is practically independent of $m_t$ and $V_{cb}$.

\begin{figure}[t]
\hspace*{3cm}\epsfig{figure=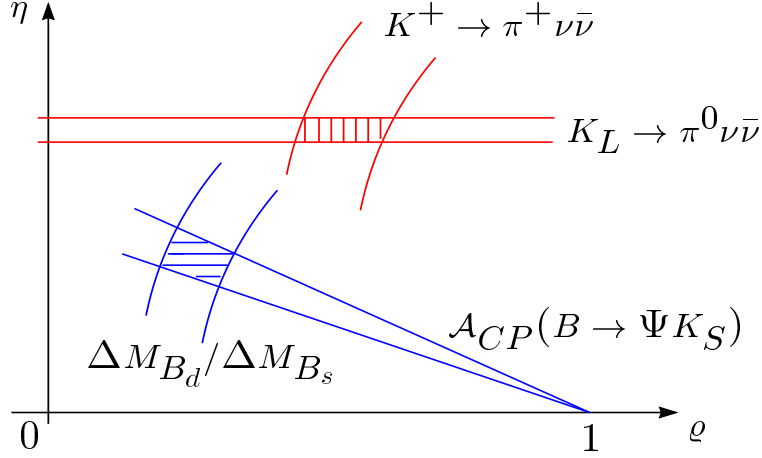,width=15cm,height=12cm}
\vspace*{-4cm}
\caption{Schematic determination of the unitarity triangle
vertex ($\varrho$, $\eta$) from the $B$ system (horizontally hatched)
and from $K\to\pi\nu\bar\nu$ (vertically hatched). Both determinations
can be performed with small theoretical uncertainty and any
discrepancy between them would indicate new physics, as illustrated
in this hypothetical example.}
\label{fig:utkb}
\end{figure}

A generic comparison between clean determinations of the unitarity
triangle in $K$ and $B$ physics is illustrated in Fig.~\ref{fig:utkb}.
Any discrepancy between the various observables would point to
new physics. In this way these rare kaon decays provide us with
an additional approach in our search for physics beyond the
SM, complementary to the direct production of new
particles. It is even possible that these rare processes lead to the
first evidence of new physics. But also in the longer run, after
new physics has already been discovered, these decays will play
an important role in analyzing in greater detail the 
underlying new dynamics.
 
\begin{figure}[h]
\hspace*{2.5cm}\epsfig{figure=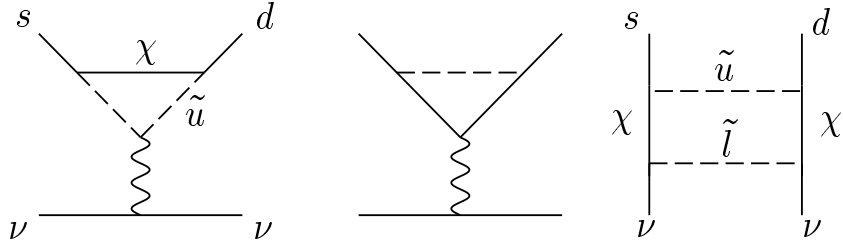,width=11.5cm,height=3.5cm}
\caption{Typical contributions to
           $K\to\pi\nu\bar\nu$ in supersymmetry with the exchange
        of squarks and gauginos.}
\label{fig:kpnnsusy}
\end{figure}

New physics contributions in $K_L\to\pi^0\nu\bar\nu$ and
$K^+\to\pi^+\nu\bar\nu$ can be parametrized in a model-independent way
by two parameters \cite{BRS}, which quantify the violation of the
golden relation (\ref{bkrel}). New effects in supersymmetric models
\cite{NW} are induced through box and penguin diagrams with new
internal particles such as charged Higgs or charginos and stops, as seen
in Fig.~\ref{fig:kpnnsusy},
replacing the $W$ boson and the 
up-type quark of the SM shown in Fig.~\ref{fig:kpnn}.
In the so-called `constrained' minimal supersymmetric standard
model (MSSM), where all flavour-changing effects are induced by
contributions proportional to the CKM mixing angles, the
`golden relation' (\ref{bkrel}) is valid. Therefore the
measurement of $B(K_L\to\pi^0\nu\bar\nu)$ and $B(K^+\to\pi^+\nu\bar\nu)$
still directly determines the angle $\beta$ and a significant violation
of (\ref{bkrel}) would rule out this model (a recent general discussion
of models with minimal flavour violation can be found in \cite{BF01}).

Given the present experimental status of supersymmetry, however, a
model-independent analysis including also flavour mixing via the
squark mass matrices is more suitable.
The new sources of flavour violation can then be parametrized by the
so-called mass-insertion approximation, in an expansion of the squark 
mass matrices around their diagonals. It turns out that
supersymmtric contributions
in this more general setting, also called the `unconstrained' MSSM,
allow for a significant violation of, for example, relation (\ref{bkrel}).
An enhancement of the branching ratios by an order of magnitude
for $K_L\to\pi^0\nu\bar\nu$ and by about a factor of three
for $K^+\to\pi^+\nu\bar\nu$ relative to the SM expectations
is possible, mostly due to the chargino-induced $Z$-penguin contribution
\cite{CI}. Recent analyses \cite{CI,BS,BCIRS} within the unconstrained
MSSM focused on the correlation of rare decays and
$\varepsilon'/\varepsilon$, the parameter of direct CP violation
in $K\to\pi\pi$ decays. This led to typical upper bounds of
$B(K_L\to\pi^0\nu\bar\nu)\leq 1.2\times 10^{-10}$ and 
$B(K^+\to\pi^+\nu\bar\nu)\leq 1.7\times 10^{-10}$.

\section{PROBING CP VIOLATION WITH $K_L\to\pi^0 e^+e^-$}

{\it G. D'Ambrosio, G. Isidori}

\noindent
The $K_L \to \pi^0 e^+ e^-$ mode is an ideal 
complement to $K_L \to \pi^0 \nu \bar{\nu}$
in clarifying the short-distance mechanism 
of CP violation and in searching for new physics. 
Within the SM both decays could provide
precise information on the CKM factor 
${\rm Im}(V_{ts}^*V_{td})$.  However, the experimental 
difficulties in reaching this goal are 
rather different in the two cases, and the two 
channels are potentially affected by different 
new-physics effects.

\subsection{Rate measurements}
Similarly to $K_L \to \pi^0 \nu \bar{\nu}$,
the direct CP-violating amplitude 
for $K_L \to \pi^0 e^+ e^-$ is dominated 
by short-distance dynamics,
i.e., by top-quark loops, and is 
calculable with high accuracy in 
perturbation theory \cite{GW,BLMM}.
Within the SM, this theoretically clean 
part of the amplitude leads to \cite{BLMM}
\begin{equation}
B(K_L\to\pi^0 e^+e^-)^{\rm SM}_{\rm CPV-dir}~=
~(2.5 \pm 0.2) \times 10^{-12}~\left[ \frac{{\rm Im}
(V_{ts}^*V_{td}) }{  10^{-4} } \right]^2, 
\end{equation}
setting to $10^{-12}$ the minimum level of sensitivity 
needed to access the short-distance information provided by
this transition.
However, contrary to $K_L \to \pi^0 \nu \bar{\nu}$,
the direct CP-violating amplitude
is not the only relevant contribution to
$K_L \to \pi^0 e^+ e^-$.
In this case it is necessary also to 
take into account the 
indirect CP-violating and the 
CP-conserving amplitudes, which 
are both dominated by long-distance 
dynamics.

The indirect CP-violating contribution alone 
can be written as 
\begin{eqnarray}
B(K_L \to \pi^0 e^+ e^-)_{\rm CPV-ind} 
&=& { \tau_L \over \tau_S } \left|\varepsilon \right|^2~
B(K_S \to \pi^0 e^+e^-) \nonumber \\ 
& =  & 3.0 \times 10^{-3}~B(K_S \to \pi^0 e^+ e^-)
\end{eqnarray}
and can therefore be controlled precisely by means of 
the $K_S \to \pi^0 e^+ e^-$ branching ratio, which is 
expected to be in the $10^{-10} - 10^{-8}$ range \cite{EPR,DEIP}.
The two CP-violating amplitudes of $K_L \to \pi^0 e^+ e^-$,
namely the direct and the indirect ones, will in general 
interfere, leading to a total CP-violating 
branching ratio that could easily reach 
the $10^{-11}$ level within the SM. 
Interestingly, the relative phase of the two amplitudes 
is determined by arg$(\varepsilon)$ and is known up to a sign. 
Thus from a measurement of both 
$B(K_S \to \pi^0 e^+e^-)$ and $B(K_L \to \pi^0 e^+ e^-)_{\rm CPV-tot}$ 
it is possible to determine ${\rm Im}(V_{ts}^*V_{td})$
up to a four-fold discrete ambiguity. 

The last component of the $K_L \to \pi^0 e^+ e^-$
amplitude is the CP-conserving (CPC) term generated by 
the process $K_L \to \pi^0 \gamma^* \gamma^* \to \pi^0 e^+ e^-$ 
\cite{KLgg,DG}. This contribution can be estimated theoretically
using experimental information on
the $K_L \to \pi^0 \gamma \gamma$ spectrum at small $M_{\gamma\gamma}$:
the data presently available from KTeV \cite{KTeV_KLpgg}
and NA48 \cite{NA48_osaka} suggest that it 
is substantially 
smaller than the direct-CP-violating component, being 
around $(1-2)\times 10^{-12}$ \cite{BDI}. 
Moreover, CP-conserving and  CP-violating 
contributions to $K_L\to\pi^0 e^+e^-$
do not interfere in the total rate and could 
be efficiently disentangled by a Dalitz-plot analysis 
\cite{BDI}. In view of these 
arguments, the CP-conserving contamination
should not represent a serious problem for  
the extraction of the interesting CP-violating 
component of the $K_L\to\pi^0 e^+e^-$ rate. 

The most serious problem in trying to 
measure $B(K_L \to \pi^0 e^+ e^-)_{\rm CPV}$ is the 
large irreducible background generated
by the process $K_L \to \gamma \gamma e^+ e^-$
\cite{Greenlee}. Imposing 
the cut $|M_{\gamma \gamma} -M_{\pi^0}|<5$~MeV
on the two-photon invariant mass spectrum 
of $K_L \to \gamma \gamma e^+ e^-$, the latter 
turns out to have
a branching ratio $\sim 3 \times 10^{-8}$,
more than $10^3$ times larger than the signal.
Employing additional cuts on various 
kinematical variables, it is possible to
reduce  this background 
down to the  $10^{-10}$ level  
\cite{Greenlee,KTeV_p0ee}, but it is hard 
to reduce it below this figure without 
drastic reductions of the signal efficiency.
We stress, however, that this does not 
imply that the signal is unmeasurable 
in a high-statistics experiment, where 
the physical background can be measured 
and modelled with high accuracy.
For instance, assuming an effective 
background level of $10^{-10}$,
one could still determine ${\rm Im}(V_{ts}^*V_{td})$
with a 10\% statistical error by collecting 
a total of $2 \times 10^{4}$ signal and background events,
i.e., with a total number of $K_L$'s  larger 
than $2 \times 10^{14}/\epsilon_{\pi^0ee}$,
where $\epsilon_{\pi^0ee}$ is the overall signal efficiency.

\subsection{Time-dependent $K_{L,S}\to\pi^0 e^+e^-$ interference}

\begin{figure}[t]
\begin{center}
\leavevmode
\epsfysize=8cm
\epsfxsize=12cm\epsfbox{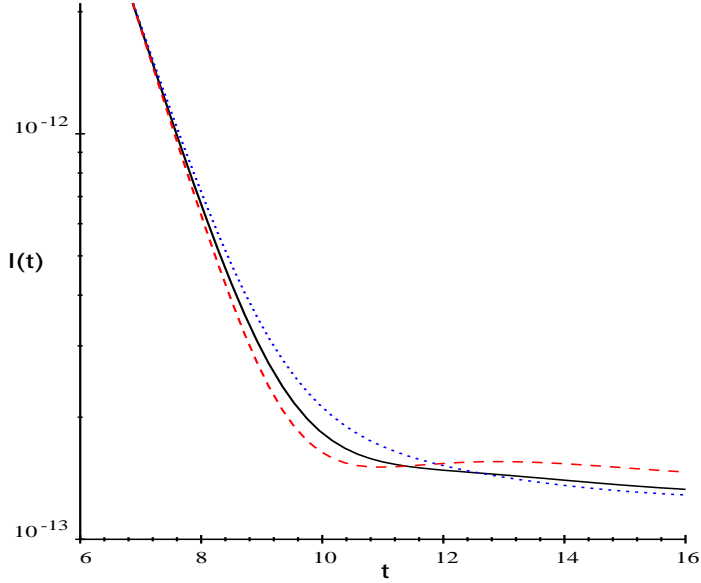}
\end{center}
\caption{Probability distribution for a $|K^0\rangle$ state
at $t=0$ to decay
into $| \pi^0(\gamma \gamma) e^+e^-\rangle$ as a function of 
time (in units of $\tau_S$), for ${\rm Im}(V_{ts}^*V_{td})=0$, 
$\pm 1.3 \times 10^{-4}$ (see text).}
\label{fig:interf}
\end{figure}

Complementary information on the direct CP-violating 
component of the $K_L\to\pi^0 e^+e^-$ amplitude 
can be obtained by studying the time evolution of the 
$K_{L,S} \to \pi^0 e^+e^-$ decay \cite{DG,timeevol}.
Although challenging from the experimental point of 
view, this method has two intrinsic advantages: 
i) the interference between $K_S$ and $K_L$ amplitudes
is only due to the CP-violating part of the latter; and
ii) the process $K_S \to \gamma \gamma e^+e^- $ is very 
suppressed with respect to $K_S \to \pi^0 e^+e^-$
[$B(K_S \to \gamma \gamma e^+e^-) \sim {\rm few} \times 10^{-12}$],
so the background due to the $| \gamma \gamma e^+e^- \rangle$
final state is almost negligible 
at small times ($t \ll \tau_L$). 

As an example, in Fig.~\ref{fig:interf} 
we show the time evolution of a pure $|K^0\rangle$ 
beam at $t=0$. In this case the 
probability distribution of decays into the final state 
$|\pi^0 e^+e^- \rangle$  or in the background channel 
$|e^+e^- \gamma \gamma \rangle$ (with $M_{\gamma \gamma} \sim M_{\pi^0}$),
as a function of the proper time $t$, 
can be written as
\begin{eqnarray}
I(t) &=& \frac{\tau_S}{2} \left\{ \left| {\cal A}_S \right|^2 e^{-t/\tau_S}
+2 {\rm Re}\left[{\cal A}_L^{\rm CPV} {\cal A}_S^* e^{-i(m_L-m_S)t} 
  \right] 
e^{-t(\tau_L+\tau_S)/2\tau_L\tau_S} \frac{}{} \right. \nonumber\\ 
&& \qquad\qquad \left. 
+ \left[ \left| {\cal A}_L^{\rm CPV} \right|^2 + 
\left| {\cal A}_L^{\rm CPC} \right|^2 + \left| {\cal A}_L^{\rm bkg.} 
   \right|^2 \right]  
e^{-t/\tau_L } \right\}~.
\end{eqnarray}
The three curves in Fig.~\ref{fig:interf}
have been obtained assuming 
$\tau_S\, |{\cal A}_S |^2 = B(K_S\to\pi^0 e^+e^-)  = 10^{-8}$ and 
$\tau_L\, |{\cal A}_L^{\rm bkg.}|^2 = B(K_L\to\gamma\gamma e^+e^-)_{\rm cuts} 
= 10^{-10}$, and employing the 
following  three values of ${\rm Im}(V_{ts}^*V_{td})$:
$0$,  $\pm 1.3 \times 10^{-4}$. 
As can clearly be seen, the interference term is 
quite sensitive to the value of the direct CP-violating amplitude.
On a purely statistical level, in this example one could reach a 10\% 
error on ${\rm Im}(V_{ts}^*V_{td})$ 
with an initial flux of $2 \times 10^{15}/\epsilon^\prime_{\pi^0ee}$
$K_L$'s, where $\epsilon^\prime_{\pi^0ee}$ denotes the 
efficiency for decays occurring within the first 15 $K_S$
decay lengths.

It is worth stressing that the unknown value of 
$B(K_S\to\pi^0 e^+e^-)$ plays a major role 
in the time distribution. If $B(K_S\to\pi^0 e^+e^-)$ is 
found to be below $10^{-9}$, then it is probably too difficult 
to measure the interference term with high precision. 
On the other hand, if $B(K_S\to\pi^0 e^+e^-) > 10^{-9}$, 
then it is worthwhile to plan a measurement of the 
$K_{L,S} \to \pi^0 e^+e^-$ time interference, not only for 
its own interest, but also to obtain a precise determination 
of $B(K_S\to\pi^0 e^+e^-)$, which is needed anyway to disentangle 
the large indirect CP-violating component of $B(K_L\to\pi^0 e^+e^-)$.
Interestingly, a heuristic argument based on vector-meson dominance
favours a large value of $B(K_S\to\pi^0 e^+e^-)$ \cite{BDI},
a conjecture that will soon be tested by the NA48 collaboration
\cite{NA48:KS}.

\subsection{Beyond the SM}
The direct CP-violating amplitude for
$K_L \to \pi^0 e^+ e^-$ is very sensitive
to possible extensions of the SM in the 
flavour sector and could eventually provide 
unambiguous evidence for new physics.
For instance, within supersymmetric models 
with generic flavour structures, 
$B(K_L\to\pi^0 e^+e^-)_{CPV}$ could be enhanced 
with respect to its SM value up to one order 
of magnitude \cite{CI,BCIRS}. 
This could happen either via a modification
of the pure electroweak amplitudes ($Z$ penguin and $W$ box),
present also in the $K_L \to \pi^0 \nu \bar{\nu}$ mode,
or via nonstandard effects in the  
$\gamma$-penguin amplitude, absent in the 
$K_L \to \pi^0 \nu \bar{\nu}$ case \cite{BCIRS}.
Interestingly, the two types of effects could 
be disentangled by means of 
$K_{L,S} \to \pi^0 e^+ e^-$ transitions only,
measuring both $B(K_L\to\pi^0 e^+e^-)_{\rm CP-dir}$
and the time-dependent interference \cite{BDI}. 
Indeed, the latter is sensitive only to the
vector component of the amplitude (dominated  
by the $\gamma$-penguin contribution), whereas 
the former is equally sensitive to the vector and 
axial components of the amplitude.


\section{CHARGED-LEPTON-FLAVOUR VIOLATION IN KAON DECAYS\\
IN SUPERSYMMETRIC THEORIES} 

{\it A. Belyaev, M. Chizhov, A. Dorokhov, J. Ellis, M. E. G\'omez,
S. Lola}

\subsection{Introductory remarks}

In this section
we discuss rare kaon decays that violate charged-lepton flavour
conservation in supersymmetric theories with and without $R$~parity:
for details and a complete set of references, see \cite{rarekaon}. 
Recent data from the Super-Kamiokande~\cite{SuKa} 
and other experiments have triggered an upsurge 
of interest in extensions
of the SM with massive neutrinos and/or violation of the 
charged-lepton numbers in
processes such as  $\mu \rightarrow e \gamma$, $\mu \rightarrow 3 e$, 
$\tau \rightarrow \mu
\gamma$ and $\mu \to e$ conversion on heavy 
nuclei~\cite{neutrinoLFVns,neutrinoLFVs,a,GELLN,KO}.
Here we discuss the prospects for muon-number violation in rare kaon
decays, encouraged by hopes that the proton driver for a neutrino
factory~\cite{mufact} could be used to improve the limits 
significantly. Any observable
rate for processes like $K^0\to \ell^+ \ell'^-$, 
$K^0\to \pi^0\ell^+ \ell'^-$, 
$K^+\to \pi^+\ell^+\ell'^-$ ($\ell\not=\ell'$), or  
$K^+\to \pi^-\ell^+ \ell'^+$ would constitute unambiguous evidence 
for new physics. The
rates for such processes remain extremely suppressed if we extend 
the SM minimally to
include right-handed neutrinos, but larger rates are possible in more 
ambitious extensions of the
SM. Supersymmetry is one example of new physics that could 
amplify rates for some of
the rare processes,  either in the minimal supersymmetric extension of the 
standard model (MSSM)
or in its modification to include the violation of $R$ parity.

\subsection{Kaon decays violating charged lepton number in the MSSM}

In this paragraph, we evaluate the
rates for lepton-number-violating
rare kaon decays in the MSSM with massive neutrinos,
assuming the seesaw mechanism~\cite{seesaw}, which we consider to be the
most natural way to obtain neutrino masses
in the sub-eV range. In particular, 
we assume Dirac neutrino masses $m_{\nu}^D$ of the same order as
the charged-lepton and quark masses, and heavy Majorana
masses $M_{\nu_R}$, leading to a light effective neutrino mass
matrix of the form:
$m_{eff}
=m^D_{\nu}\cdot (M_{\nu_R})^{-1}\cdot m^{D^{T}}_{\nu}$

\begin{figure*}[htb]
\hspace*{2.8 cm}
{\unitlength=1. pt
\SetScale{1.}
\SetWidth{0.7}      
{} \qquad\allowbreak
\begin{picture}(100,80)(0,0)
\ArrowLine(30,20)(0,20)
\Line(30,20)(70,20)
\ArrowLine(100,20)(70,20)
\DashLine(30,20)(30,60){1.0}
\DashLine(70,20)(70,60){1.0}
\ArrowLine(0,60)(30,60)
\Line(30,60)(70,60)
\ArrowLine(70,60)(100,60)
\Text(15.0,70.0)[r]{$s$}
\Text(55.0,70.0)[r]{$\chi^0$}
\Text(90.0,70.0)[r]{$\mu$}
\Text(15.0,10.0)[r]{$d$}
\Text(55.0,10.0)[r]{$\chi^0$}
\Text(90.0,10.0)[r]{$e$}
\Text(25.0,40.0)[r]{$\tilde{d}_i$}
\Text(80.0,40.0)[r]{$\tilde{\ell}_i$}
\end{picture} 
{} \qquad\allowbreak
\begin{picture}(100,80)(0,0)
\ArrowLine(30,20)(0,20)
\Line(30,20)(70,20)
\ArrowLine(100,20)(70,20)
\DashLine(30,20)(30,60){1.0}
\DashLine(70,20)(70,60){1.0}
\ArrowLine(0,60)(30,60)
\Line(30,60)(70,60)
\ArrowLine(70,60)(100,60)
\Text(15.0,70.0)[r]{$s$}
\Text(55.0,70.0)[r]{$\chi^\pm$}
\Text(90.0,70.0)[r]{$\mu$}
\Text(15.0,10.0)[r]{$d$}
\Text(55.0,10.0)[r]{$\chi^\pm$}
\Text(90.0,10.0)[r]{$e$}
\Text(25.0,40.0)[r]{$\tilde{u}_i$}
\Text(80.0,40.0)[r]{$\tilde{\nu}_i$}
\end{picture}
}
\caption{\label{fig1}
MSSM box diagrams for $K^0 \rightarrow \mu^\pm e^\mp$.
There is another neutralino exchange
diagram corresponding to the permutation of the $\mu$ and $e$.
Since $\chi^0$ is a Majorana spinor,
there are contributions from the neutralinos that
differ in the number of mass insertions.}
\end{figure*}
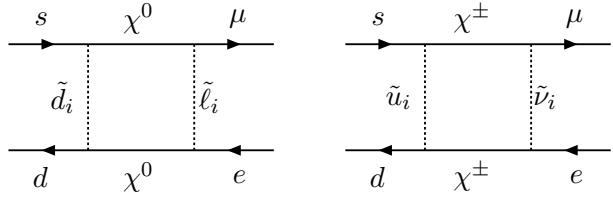

In the MSSM framework, rare kaon decays are generated by
box diagrams involving the exchanges of charginos $\chi^\pm$
and neutralinos $\chi^0$. For instance,
for $K^0 \rightarrow \mu^\pm  e^\mp$ we have the
diagrams of Fig. \ref{fig1}.

Our procedure for evaluating these contributions is as follows: we  find 
the maximal squark
mixing that is allowed by the neutral-kaon mass difference $\Delta m_K$ 
\cite{LFVK}, then  we
find the maximal slepton mixing allowed by $\mu \rightarrow e \gamma$ 
and $\mu$--$e$ conversion
in nuclei {\em in a model-independent way}. Finally, having fixed these 
values, we calculate the rates for the rare kaon decays of interest here.

We parametrize the supersymmetric  masses in terms of universal 
GUT-scale parameters $m_0$ 
and $m_{1/2}$, for sfermions and gauginos respectively, and use the 
renormalization-group
equations of the MSSM to calculate the low-energy sparticle masses. 
Other relevant free
parameters of the MSSM are the trilinear coupling $A$, for which we use
the initial
condition $A_0=- m_{1/2}$, the sign of the Higgs mixing parameter $\mu$, 
and the value of
$\tan\beta$.  Models with different signs of  $\mu$ give similar results: 
here we assume $\mu < 0$.

\begin{figure}[p]
\begin{center}
\begin{tabular}{c c}
\epsfig{file=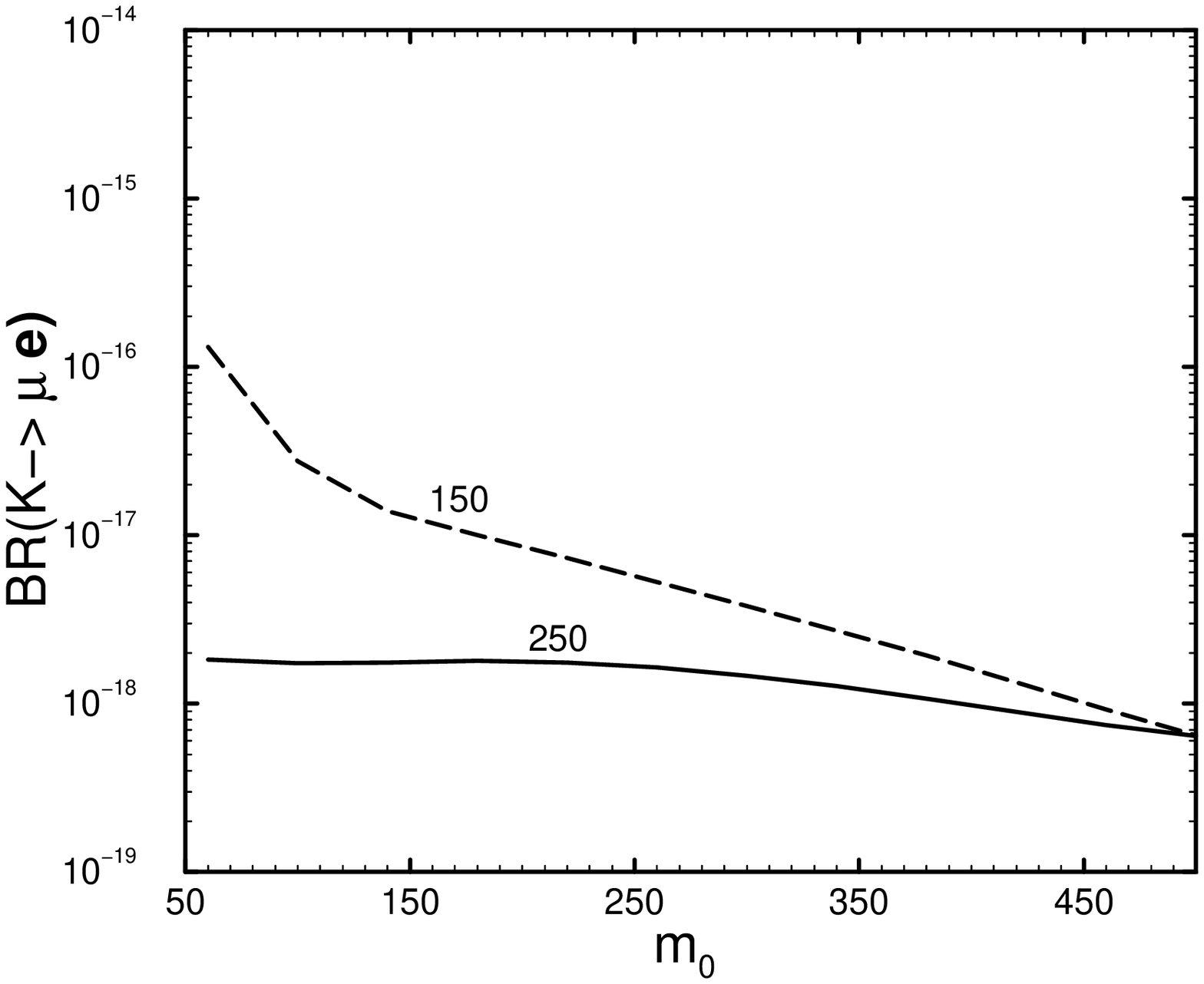, width=6.5cm,height=6.0cm} &
\epsfig{file=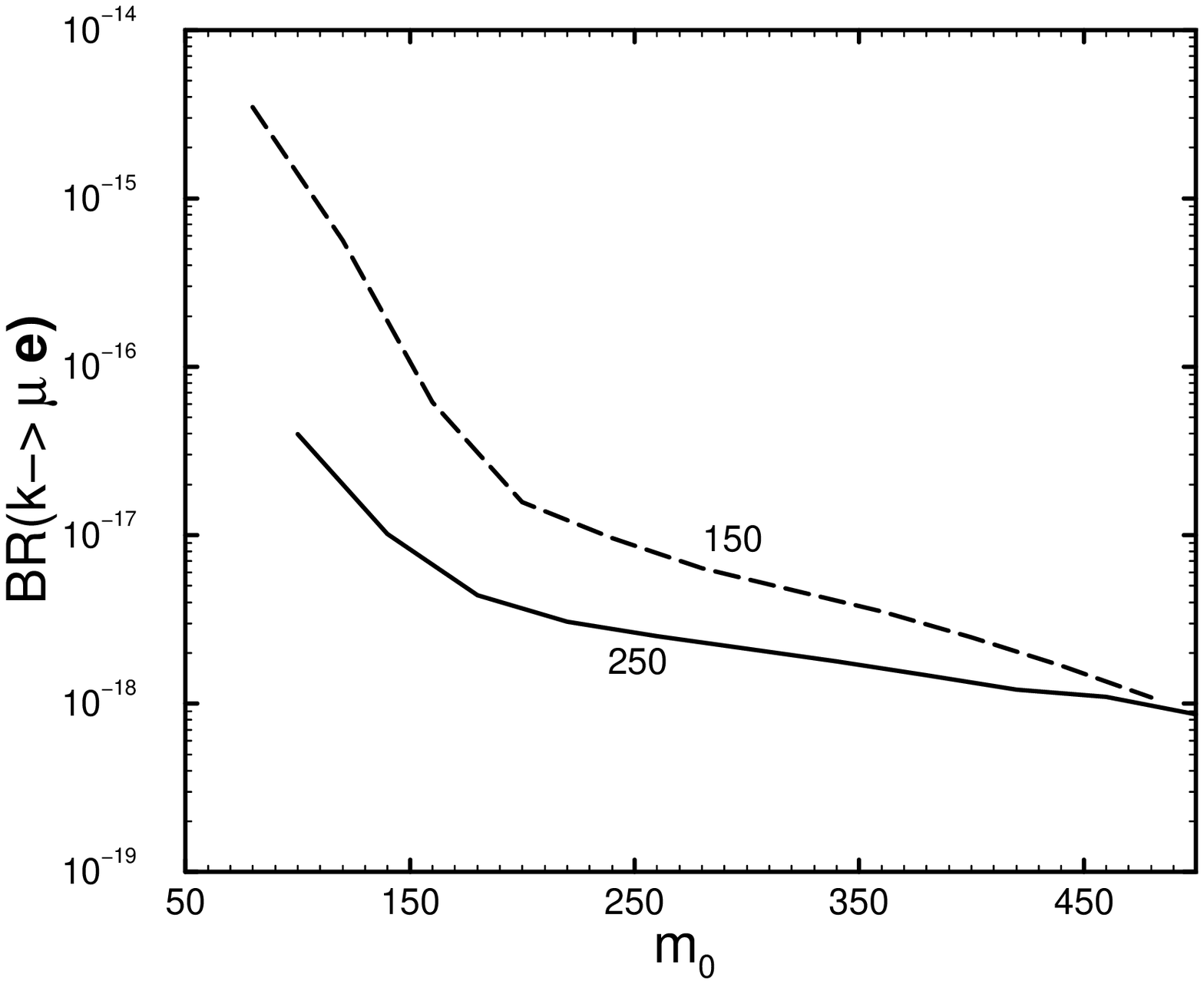, width=6.5cm,height=6.0cm} \\
\epsfig{file=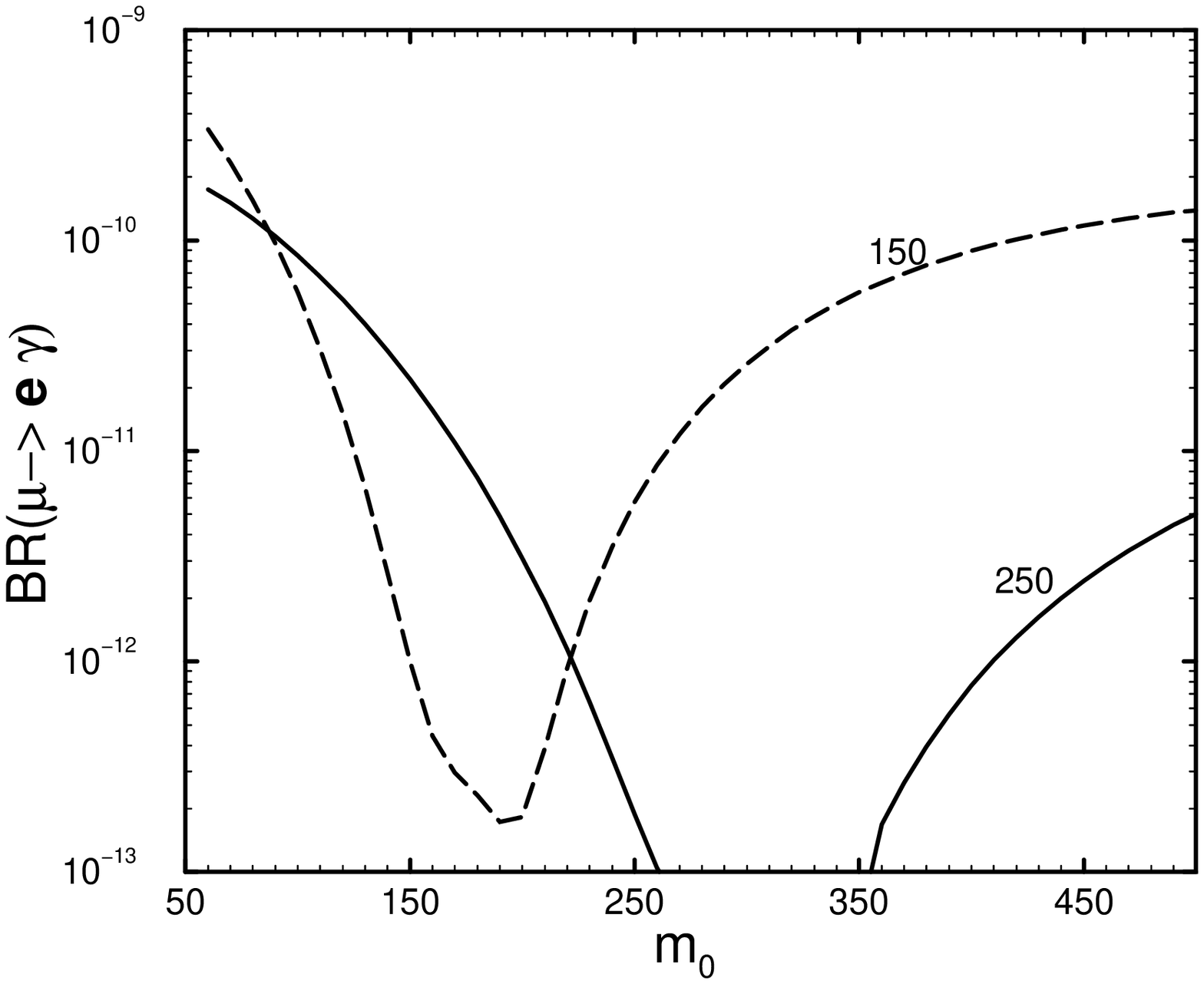, width=6.5cm,height=6.0cm} &
\epsfig{file=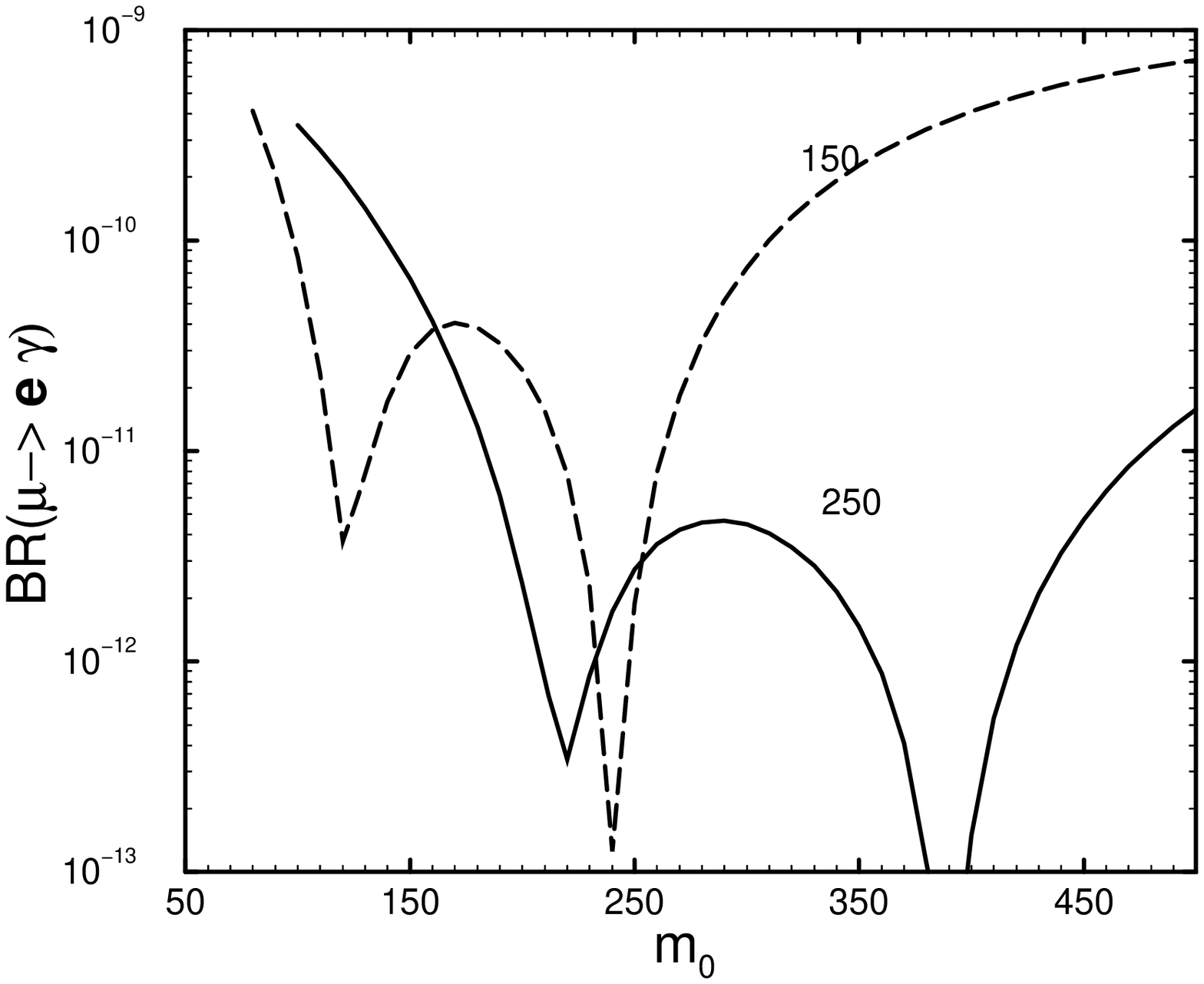, width=6.5cm,height=6.0cm} \\
\epsfig{file=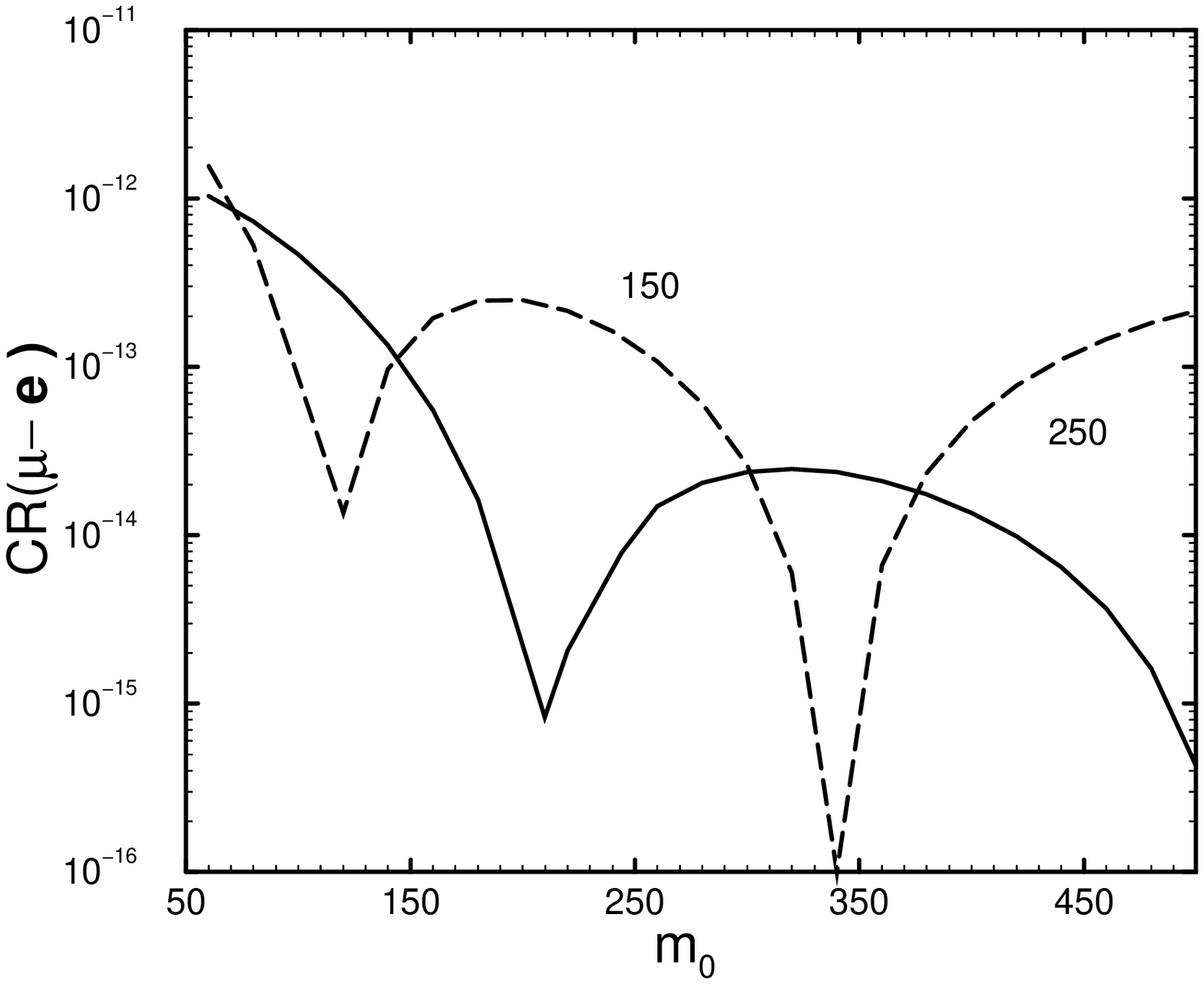, width=6.5cm,height=6.0cm} &
\epsfig{file=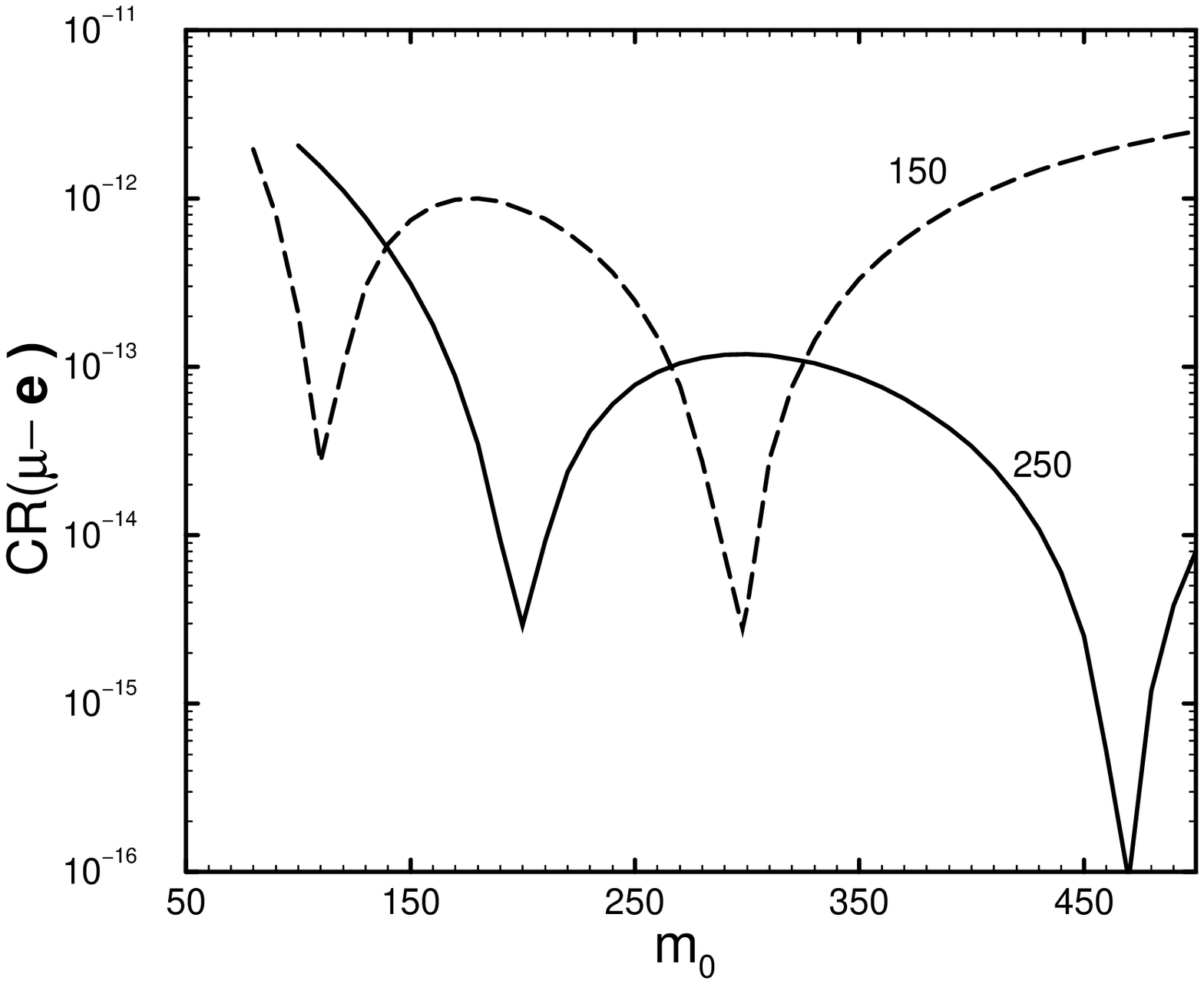, width=6.5cm,height=6.0cm} \\
\end{tabular}
\end{center}
\caption{\it Illustrative predictions for $B(K^0 \to \mu e)$, $B(\mu \to e
\gamma)$ and the $\mu \to e$ conversion rate, for $\tan\beta = 10$
(left column) and $\tan\beta = 20$ (right column), and $m_{1/2} =
150$~GeV (dashed lines) or $250$~GeV (solid lines), as functions of
$m_0$ (in GeV).}
\label{fig:BCDEGL}
\end{figure} 
Typical results are shown in Fig.~\ref{fig:BCDEGL}.
As expected, the larger the value of $\tan\beta$ and the smaller the soft 
supersymmetric terms,
the larger the branching ratios, apart from certain cancellations. In the 
case $\tan\beta = 10$
and $m_{1/2} = 250$ GeV,  for the range $m_0 \ge 170$~GeV where 
$B(\mu \to  e \gamma)$ and $R(\mu\to e)$ are consistent with the 
current experimental bounds, $B(K \to \mu e)$ is at
most $2 \times 10^{-18}$. However, for the same value of $m_{1/2}$,  when 
$\tan\beta = 20$ we
find a significantly larger branching ratio at small values of 
$m_0 \sim 170$~GeV. Moreover, for
smaller $m_{1/2} = 150$~GeV, we gain almost two orders of magnitude when 
we consider $m_{0}$ in
the low-mass window between 100 and 150 GeV. We recall~\cite{GELLN} that 
these lower  values of
$m_{1/2}, m_0$ are consistent with accelerator constraints and generically 
yield cold dark
matter densities in the range prefered by cosmology~\cite{EFGO}.

This analysis demonstrates that, despite the limits from 
$\mu \rightarrow e \gamma$,
$\mu$--$e$ conversion  and $\Delta m_K$, the branching ratio of 
$K \rightarrow \mu e$  may be
within the reach of the next generation of experiments,  namely in the 
range $10^{-16}
\rightarrow 10^{-18}$, at least if $\tan\beta$ is large and the soft 
supersymmetry-breaking
terms are small. 

\subsection{Kaon decays violating charged lepton number in
$R$-violating supersymmetry}

We now discuss kaon decays violating 
charged-lepton flavour beyond the context of the MSSM.
As is well known, the gauge symmetries of the 
MSSM allow additional dimension-four
Yukawa couplings, of the form
$
\lambda L_{i}L_{j}{\bar{E}}_{k}, \; \;
\lambda ^{\prime }L_{i}Q_{j}{\bar{D}_{k}}, \; \;
\lambda ^{\prime \prime }{\bar{U}_{i}}{\bar{D}_{j}}{\bar{D}_{k}} 
$
where the $L(Q)$ are the left-handed lepton (quark) superfields, and the
${\bar{E}}$,(${\bar{D}},{\bar{U}}$) are the corresponding right-handed fields. 
If all these couplings were 
present simultaneously in the low-energy Lagrangian, they would generate
unacceptably fast proton decay. 
In this study, we allow only lepton-number-violating processes and study  
the general case
when   several $R$-violating operators may be non-zero. Then we discuss the 
limits on their
combinations that are obtainable from kaon decays.

What is the connection of these limits with neutrino masses? We note that 
neutrino masses do
not strictly constrain $K \rightarrow \mu e$ (and in certain cases the rest 
of the
flavour-violating-processes), since neutrino masses may only constrain 
products of
$LL\bar{E}$ or $LQ\bar{D}$ operators, but not mixed $LL\bar{E}$-$LQ\bar{D}$ 
products. Even
for the diagrams with products of only $LQ\bar{D}$ operators, rare kaon 
decays involve quarks
of the lightest and second-lightest generations. In this case the bounds 
from neutrino masses
are significantly  weaker, and the stricter limits come from the current 
measurements of the
rare kaon decays themselves.

\begin{figure}[h]
\hspace*{1.8 cm}
{\unitlength=0.7 pt
\SetScale{0.7}
\begin{picture}(220,100)(0,-10)
\SetWidth{1.}      
\ArrowLine(40,35)(15,35)    \ArrowLine(15,45)(40,45)   \Text(30,55)[]{$K^0$}
\ArrowArcn(110,5 )(80,150,25) \ArrowArcn(110,75)(80,-25,-150) 
\Text(50,46)[]{$d$}
\Text(50,34)[]{$\bar s$}
\Vertex(180,40){3}
\DashLine(180,40)(210,40){5}
\ArrowLine(240,0)(210,40) \Text(245,0)[l]{$\ell^+$}
\ArrowLine(210,40)(240,80)\Text(245,80)[l]{$\ell^-$}
\Vertex(210,40){3}
\Text(190,50)[]{$\tilde\nu$}
\end{picture}

\vspace*{-2.5cm}
\hspace*{8.7cm}
\begin{picture}(220,100)(0,-10)
\SetWidth{1.}      
\ArrowLine(40,35)(15,35)    \ArrowLine(15,45)(40,45)   \Text(30,55)[]{$K^0$}
\ArrowArcn(110,5 )(80,150,55) \ArrowArcn(110,75)(80,-55,-150) 
\Text(50,46)[]{$d$}
\Text(50,34)[]{$\bar s$}
\DashLine(160,15)(160,65){5} 
\Text(165,40)[l]{$\tilde u$}
\Vertex(160,10){3} \Vertex(160,70){3}
\ArrowLine(190,10)(160,10) \Text(195,10)[l]{$\ell^+$}
\ArrowLine(160,70)(190,70) \Text(195,70)[l]{$\ell^-$}
\end{picture}
}
\caption{\label{fig:kaon0}
Diagrams 
involving $R$-violating couplings that
yield two-body 
$K^0 \rightarrow \ell^\pm \ell^\mp$ decays.}
\end{figure}
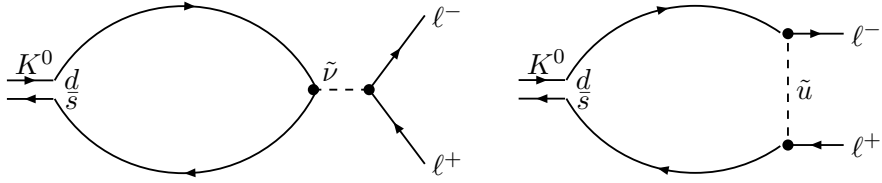

$\bullet$
Two-body $K^0$ decays to muons and electrons proceed via the diagrams 
shown in
Fig.~\ref{fig:kaon0}. 
Using the Feynman rules derived for the relevant 
effective kaon, pion and
lepton interactions, we have  recalculated 
the important kaon
decay processes, and update the limits on the products of  $R$-violating 
couplings using the
present experimental limits.
The diagrams of Fig.~\ref{fig:kaon0} lead to the 
following effective Lagrangian for $K^0 \ell^+ \ell^-$ interactions: 
\begin{eqnarray}
{\cal L}_{K^0  \ell^-_j \ell^+_k} &=&
{F_{K^0}\over 2 m_{\tilde\nu_i}^2}\left[
\lambda^*_{ijk} \lambda'_{i12}\left(\overline{{\ell_j}_L}{\ell_k}_R\right)-
\lambda_{ikj} \lambda'^*_{i21}\left(\overline{{\ell_j}_R}{\ell_k}_L\right)
\right]K^0(p_K) \nonumber \\
 &-&{f_K\over 4 m_{\tilde u_i}^2} \lambda'^*_{ji1}\lambda'_{ki2}
~p_K^\mu \left(\overline{{\ell_j}_L}\gamma_\mu {\ell_k}_L\right)K^0(p_K)\ ,
\end{eqnarray}
where $F_{K^0}=m_{K^0}^2 f_{K}/(m_s+m_d)$,  $m_s+m_d \simeq 0.15$ GeV is the 
sum of the current
masses of the $s$ and $d$ quarks, and $f_K=0.1598$ GeV is the kaon decay 
constant. The value of
$F_{K^0}$ is related to the pseudoscalar 
$\langle 0|\bar s\gamma^5 d|K^0\rangle =-F_{K^0}$ 
matrix element, and
is obtained from $f_K$ by  using the Dirac equations for quarks.  All QCD 
corrections are
included in this phenomenological approach. In the following, we assume 
that the $R$-violating
couplings are real and that only one of their products is non-zero.

We have implemented the Feynman rules in the CompHEP package~\cite{comphep}, 
using the effective
Lagrangian, and have obtained analytical results for the kaon decay 
width as well
as limits  on the products of $\lambda\lambda'$ and $\lambda'\lambda'$ 
couplings from sneutrino
and squark exchange, respectively. The results are shown in equations (16)
and (17) of~\cite{rarekaon}.

We now discuss the diagrams for 3-body kaon
decays to pions and two charged leptons, of which
there are two qualitatively different kinds.

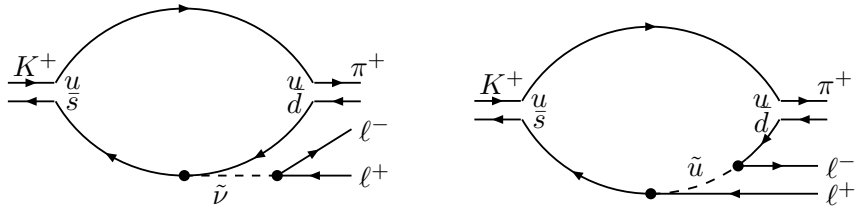
\begin{figure}[h]
\vspace*{-.7cm}
\hspace*{2.cm}
{\unitlength=0.7 pt
\SetScale{0.7}
\begin{picture}(220,110)(0,0)
\SetWidth{1.}      
\ArrowLine(40,35)(15,35)    \ArrowLine(15,45)(40,45) \Text(30,55)[]{$K^+$}
\ArrowLine(205,35)(180,35) \ArrowLine(180,45)(205,45)\Text(210,55)[]{$\pi^+$}
\ArrowArcn(110,5 )(80,150,30) \ArrowArcn(110,75)(80,-90,-150)
\ArrowArcn(110,75)(80,-30,-90)
\DashLine(110,-5)(160,-5){5}
\Vertex(110,-5){3}
\Text(130,-15)[]{$\tilde\nu$}
\ArrowLine(200,-5)(160,-5)\Vertex(160,-5){3}\Text(205,-5)[l]{$\ell^+$}
\ArrowLine(160,-5)(200,20)\Text(205,20)[l]{$\ell^-$}
\Text(50,45)[]{$u$}
\Text(50,35)[]{$\bar s$}
\Text(170,35)[]{$\bar d$}
\Text(170,45)[]{$ u$}
\end{picture}

\vspace*{-2.5cm}
\hspace*{8.3cm}
\begin{picture}(220,110)(0,0)
\SetWidth{1.}      
\ArrowLine(40,35)(15,35)    \ArrowLine(15,45)(40,45) \Text(30,55)[]{$K^+$}
\ArrowLine(205,35)(180,35) \ArrowLine(180,45)(205,45)\Text(210,55)[]{$\pi^+$}
\ArrowArcn(110,5 )(80,150,30) \ArrowArcn(110,75)(80,-90,-150) 
\ArrowArcn(110,75)(80,-30,-50) \DashCArc(110,75)(80,-90,-50){5}
\Text(130,10)[l]{$\tilde u$}
\Text(50,35)[]{$\bar s$}
\Text(50,45)[]{$u$}
\Text(170,45)[]{$ u$}
\Text(170,35)[]{$\bar d$}
\ArrowLine(200,-5)(110,-5)\Vertex(110,-5){3}\Text(205,-5)[l]{$\ell^+$}
\ArrowLine(157,10)(200,10)\Text(205,10)[l]{$\ell^-$}
\Vertex(157,10){3}
\end{picture}
}

\caption{\label{fig:kaon1}
\it Diagrams  involving $R$-violating couplings that yield the three-body
leptonic decays $K^+ \to\pi^+ \ell^{-}\ell^{+}$.}
\end{figure}

$\bullet$
The kaon may decay into a pion of the same charge, in which case the
leptons in the final state must have opposite signs:  
$K^\pm\to\pi^\pm \ell^\mp\ell'^\pm$
and $K^0\to\pi^0 \ell^\mp\ell'^\pm$. The corresponding
diagram for the first process is shown in Fig.~\ref{fig:kaon1}.
The limit obtained from $K^0\to\ell^+\ell'^-$ is typically 1-2 orders 
of magnitude better than that derived from $K^+\to \pi^+\ell^+\ell'^-$
decay, so we do not discuss further this class of three-body decays.

$\bullet$
The kaon may decay into a pion with the opposite charge, in which
case the leptons in the final state must have the same signs: 
$K^\pm\to\pi^\mp \ell^\pm\ell'^\pm$.
This process involves two heavy virtual  particles, the $W$ boson and a
down squark. One should note that the decay width of this process is 
directly proportional
to the mixing between the left- and right-handed squark states,
denoted by $\tilde b_L$  and $\tilde b_R$, respectively.
If there is no mixing, the same-sign-lepton
process vanishes. 
The effective
Lagrangian includes the following terms:
\begin{eqnarray}
{\cal L}_{K^+ \ell^-_i \bar\nu_i } &=& 
V_{us} \sqrt{2} G_F f_K p_K^\mu 
\left(\overline{{\nu_i}_L} \gamma_\mu {\ell_i}_L \right)K^+(p_K)\ , \\
{\cal L}_{\pi^+ \ell^-_i \bar\nu_i } &=& 
V_{ud} \sqrt{2} G_F f_\pi p_\pi^\mu 
\left(\overline{{\nu_i}_L} \gamma_\mu {\ell_i}_L \right)\pi^+(p_\pi)\ .
\end{eqnarray}
Here $f_{K}$ and $f_{\pi}=0.1307$ GeV 
are the kaon and pion decay constants, respectively,
$G_F=1.16639 10^{-5}$ GeV$^{-2}$ is the Fermi constant 
and  $V_{us},V_{ud} $ are CKM matrix elements. We also have
\begin{eqnarray}
{\cal L}_{K^+ (\ell^-_j)^C \nu_i } (\tilde d_k) &=& 
\frac{\lambda'_{ik2}\lambda'_{j1k} V_{LR}}{4 m_{\tilde d_k}^2}
 F_{K^+} \left(\overline{({\ell_j}_L)^C} {\nu_i}_L\right)K^+(p_K)\ , \\
{\cal L}_{\pi^+ (\ell^-_j)^C \nu_i } (\tilde d_k) &=& 
\frac{\lambda'_{ik1}\lambda'_{j1k}V_{LR}}{4 m_{\tilde d_k}^2}
F_{\pi^+}\left(\overline{({\ell_j}_L)^C} {\nu_i}_L\right)\pi^+(p_\pi)\ ,
\end{eqnarray}
where $F_{K^+}=m^2_{K^+}f_K/(m_s+m_u)$,
$F_{\pi^+}=m^2_{\pi^+}f_\pi/(m_d+m_u)$,
$m_s+m_u\simeq 0.15$ GeV, and $m_d+m_u\simeq 0.01$ GeV
and $V_{LR}$ denotes the left-right squark mixing matrix element.

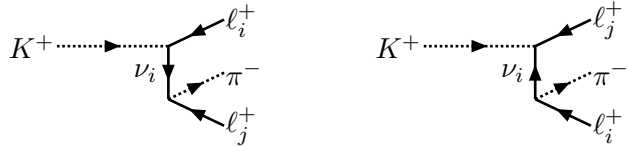
\begin{figure}[h]
\hspace*{2.8cm}
{\unitlength=1.0 pt
\SetScale{1.}
\SetWidth{1.0}      
{} \qquad\allowbreak
\begin{picture}(95,79)(0,0)
\Text(15.0,60.0)[r]{$K^+$}
\DashArrowLine(16.0,60.0)(58.0,60.0){1.0}
\Text(80.0,70.0)[l]{$\ell^+_i$}
\ArrowLine(79.0,70.0)(58.0,60.0)
\Text(54.0,50.0)[r]{$\nu_i$}
\ArrowLine(58.0,60.0)(58.0,40.0)
\Text(80.0,50.0)[l]{$\pi^-$}
\DashArrowLine(58.0,40.0)(79.0,50.0){1.0}
\Text(80.0,30.0)[l]{$\ell^+_j$}
\ArrowLine(79.0,30.0)(58.0,40.0)
\end{picture} \
{} \qquad\allowbreak \ \ \ \ \
\begin{picture}(95,79)(0,0)
\Text(15.0,60.0)[r]{$K^+$}
\DashArrowLine(16.0,60.0)(58.0,60.0){1.0}
\Text(80.0,70.0)[l]{$\ell^+_j$}
\ArrowLine(79.0,70.0)(58.0,60.0)   
\Text(54.0,50.0)[r]{$\nu_i$}
\ArrowLine(58.0,40.0)(58.0,60.0)
\Text(80.0,50.0)[l]{$\pi^-$}
\DashArrowLine(58.0,40.0)(79.0,50.0){1.0}
\Text(80.0,30.0)[l]{$\ell^+_i$}
\ArrowLine(79.0,30.0)(58.0,40.0)
\end{picture} \
}
\vspace*{-1.cm}
\caption{\label{fig:kaon3}
\em Diagrams for the like-sign lepton decay
$K^+\to\pi^- \ell^+_i\ell^+_j$,
in terms of the effective standard-model-like interactions $K \ell\nu$ and
$\pi \ell\nu$ and effective $K \ell^C\nu$ and $\pi\ell^C\nu$ interactions
related to $R$-violating operators.}
\end{figure}

The Feynman diagrams for $K^+\to\pi^- \ell^+_i\ell^+_j$ decay 
are given in terms of these effective interactions, as shown in
Fig.~\ref{fig:kaon3}.
As an example, one can obtain a constraint on the 
product of $\lambda'_{2k2}\lambda'_{11k}$ and $V_{LR}$ 
from ${K^+\to \pi^- e^+\mu^+}$ decay.
In this case, we have the following numerical result
\footnote{
In  \cite{LitShr} a similar constraint was found with a different choice
of diagrams.}:
$V_{LR}(\lambda'_{2k2}\lambda'_{11k})\times
\left(\frac{100~\mbox{GeV}}{m_{\tilde d_k}}\right)^2 \le 10.
$
It is apparent that kaon decay into a pion and a like-sign lepton pair
is too strongly suppressed to be useful at present.

\subsection{Conclusions}

We have discussed the flavour-violating decays of
kaons into charged-lepton pairs in supersymmetric theories,
in both the minimal supersymmetric standard model and
$R$-violating models. In the first case, despite the
limits from $\mu \rightarrow e \gamma$,
$\mu$--$e$ conversion 
and $\Delta m_K$, the kaon decay branching ratios
for large $\tan\beta$ and
small soft supersymmetry-breaking terms may be accessible
to a future generation of experiments using new intense proton sources.
In the case of  $R$-violating supersymmetry,  we studied the expected rates 
for the decays $K
\rightarrow \mu^\pm e^\mp$ and $K \rightarrow \pi \mu e$, for all two- 
and three-body
processes, and  obtained the bounds on products of $LL\bar{E}$ and 
$LQ\bar{D}$ operators
summarized in (16) and (17) of~\cite{rarekaon}. We have also noted the 
possibility of
like-sign lepton events in the presence of non-zero $\tilde{b}_L$ --
$\tilde{b}_R$ mixing, but
for this to occur at a significant rate one would need large $R$-violating 
couplings.
Our final conclusion is  that lepton-flavour-violating rare kaon decays 
have the potential to
provide important information on the issue of flavour physics. Any future 
observation would,
in addition, help distinguish between different supersymmetric theories.


\section{THE RARE DECAY $K^+ \to \pi^- l^+ l^+$}

{\it K. Zuber}

\noindent
The decays $K^+ \to \pi^- l^+ l^+$ can also be seen in the 
broader context
of $\Delta L = 2$ lepton-number-violating processes. They are
analogous to
neutrinoless double-beta decay, and ways to determine three out of nine 
matrix elements 
of effective Majorana masses $\langle m_{\alpha \beta} \rangle$
\begin{equation}\label{meffmatrix}
  \langle m_{\alpha \beta}\rangle = \left| \displaystyle 
         \sum m_m \eta^{\rm CP}_m U_{\alpha m} U_{\beta m} \right|
   \mbox{ with }  \alpha, \beta = e , \, \mu  , \, \tau .     
\end{equation}
with the relative CP-phases $\eta^{\rm CP}=\pm 1$.
For a general overview of $\Delta L = 2$ processes studied 
within this context, see \cite{zub00a}.
The current limits on the possible processes involving kaons are given by 
the BNL E865 experiment \cite{BNL865b}
\begin{eqnarray}
B(K^+ \to \pi^- e^+ e^+) < 6.4 \times 10^{-10} \\
B(K^+ \to \pi^- \mu^+ e^+) < 5.0 \times 10^{-10} \\
B(K^+ \to \pi^- \mu^+ \mu^+) < 3.0 \times 10^{-9} 
\label{mumulimit}
\end{eqnarray}
However, neutrinoless double beta decay and $\mu - e$ conversion on nuclei 
are already much
more sensitive than the first two decays, and the main interest would be 
the decay $K^+ \to \pi^- \mu^+ \mu^+$ measuring  
\begin{equation}
\langle m_{\mu \mu} \rangle = \mid \sum U_{\mu m}^2 m_m \eta^{\rm CP}_m \mid 
\end{equation}
Whilst the experimental limit (\ref{mumulimit}) is not yet good enough to
restrict 
$\langle m_{\mu \mu} \rangle $ and the combination
$\lambda^{\prime}_{211}\lambda^{\prime}_{212}$ of $R$-violating
supersymmetric couplings (see sec. 1.4.3 and \cite{LitShr}), it does
lead to an improvement by about two orders of magnitude in the bound on
the mixing of 
muon neutrinos with heavy neutrinos in the mass region of 
249 MeV $< m_H <$ 385 MeV due to resonance effects \cite{dib}.

\vspace*{1cm}



\section{RARE KAON DECAY EXPERIMENTS}

{\it A. Ceccucci}

\subsection{Introduction}

We focus on the rare kaon decays which can drive an experimental
programme. 
Therefore the experimental status and the perspectives for future
measurements of the 
following decay modes are reviewed:
\begin{itemize} 
\item{$ K^+ \rightarrow \pi^+ \nu \bar{\nu}$}
\item{$K_L \rightarrow \pi^0 \nu \bar{\nu}$}
\item{$K_L \rightarrow \pi^0 l^+ l^-$}
\item{$K_L \rightarrow \mu e$}
\end{itemize}
The Standard Model (SM) theoretical predictions and the current
experimental 
limits for the above reactions are shown in Table 2.
It is very likely that the availability of high-intensity kaon 
beams will trigger other experimental proposals, ranging from the 
search for CPT violation in kaon decays
to the study of hypernuclei, but we consider the above reactions to be the
most likely physics drivers for such a programme.     

\newpage

\begin{center}
{\small Table 2: Comparison between theory and experiment.
Experimental upper limits are at 90\% CL.}
\vskip0.2cm
\begin{tabular}{|l|c|c|c|l|}
\hline
Decay Mode & SM & Exp. Result & Experiment & Technique \\
\hline
$ K^+ \rightarrow \pi^+ \nu \bar{\nu}$ & $1\times 10^{-10}$ 
&$(1.5^{+3.4}_{-1.2})\times 10^{-10}$ &E787 & kaon decay at rest \\
\hline
$K_L \rightarrow \pi^0 \nu \bar{\nu}$ & $3\times10^{-11}$  & 
$< 5.9\times 10^{-7}$& KTEV &$(\pi^0\rightarrow e^+e^-\gamma)$ \\
& &  $< 1.6\times 10^{-6}$& KTEV &  $(\pi^0\rightarrow \gamma \gamma)$ \\

& & $<2.6\times 10^{-9} $ &E787 & \begin{minipage}{35mm}{Model indep.
limit \\ using $K^+ \rightarrow \pi^+ \nu \bar{\nu}$}\end{minipage}  \\
\hline
 $K_L \rightarrow \pi^0 e^+ e^-$ & $5\times 10^{-12}$ &
$< 5.1\times10^{-10}$ 
& KTEV & \\
\hline 
$K_L \rightarrow \pi^0 \mu^+ \mu^-$ & $1\times 10^{-12}$ & 
$< 3.8\times10^{-10}$ & KTEV & \\
\hline
$K_L \rightarrow \mu e$& -- &$<4.7 \times 10^{-12}$ & E871& double 
spectrometer \\
\hline  
\end{tabular}
\end{center}

\subsection{$ K^+ \rightarrow \pi^+ \nu \bar{\nu}$}

The one event measured by AGS-E787 was observed in the 1995-1997 data 
sample, with an estimated background from all sources of
$0.08 \pm 0.02$ events ~\cite{ADL}. 
The event was obtained from data in the phase-space region above the 
$K^+ \rightarrow \pi^+ \pi^0~(K_{\pi 2})$ peak. E787 also plans to access
the 
region below the $K_{\pi 2}$ peak, where a Signal-to-Noise (S/N) 
ratio of about unity is expected for the SM prediction. 
The two most significant backgrounds come from $K_{\pi 2}$ and
$K^+ \rightarrow \mu^+ \nu~ (K_{\mu 2})$. Other important 
backgrounds derive from the $K^+$ charge-exchange reaction (CEX) and from 
$\pi^+$ particles in the beam that scatter in the detector. 
The kaons are stopped in a scintillating-fibre target in the centre 
of the detector, and a low-mass central drift chamber measures the
momenta of 
the decay products. The energy, range, and the decay sequence of charged
particles are measured in the range-stack array of scintillators. Within
the range stack, tracking is provided by two layers of straw tubes. 
The photon-veto system covers the entire solid angle. 

Incremental improvement over E787 will be provided by
AGS-E949~\cite{E949},
which will be the primary user of the AGS in the years
2001-2003.
About $3\times 10^6 K^+$ at 700 MeV/{\it c} with a $K/\pi$ ratio of
four are produced for $10^{13}$ incident protons.    
The improvement in sensitivity per year over E787 is a factor of 13. 
After two years of data taking, E949 should reach a sensitivity of 
$1.7 \times 10^{-11}$. 
Including the acceptance below the $K_{\pi 2}$ peak, one would 
expect 7 -- 13 events if the branching ratio is that predicted by the SM.
 
Further improvement in the measurement of 
$K^+\rightarrow\pi^+\nu\bar{\nu}$ is expected from the 
CKM experiment at the Fermilab Main Injector~\cite{CKM}, which
plans to exploit a decay-in-flight technique. 
The critical component is an intense (30 MHz) RF-separated 
22 GeV/{\it c} $K^+$ beam based upon 3.9 GHz transverse mode
super-conducting RF cavities.
The experiment aims for a kaon-to-pion ratio of two for decays in the 
fiducial volume. 
The separator is formed by two RF cavities operating in a mode which 
deflects the beam laterally.   
The separation between the two cavities is chosen to deflect both $\pi^+$
and protons back onto the optical axis, where they are blocked by a beam
stopper.
The charged particles in the reaction are measured by a pair of Ring
Imaging
Cherenkov counters (RICH) and independently by upstream and downstream 
magnetic spectrometers. 
The veto system has a component surrounding the decay volume and a forward
component after the $\pi^+$ RICH. The inefficiency to veto at least one
photon
from a $\pi^0$ should be $< 10^{-7}$.
A muon veto follows the spectrometer to control the $K_{\mu 2}$ and 
$K^+ \rightarrow \mu^+ \gamma \nu$ backgrounds, and the inefficiency to
detect muons is required to be $< 10^{-5}$. 
The Fermilab Main Injector has now been in operation for over a
year, and
the beam line to transport 120 GeV slow-spill protons is under
construction.
This beam should be available for slow-spill tests in 2002. 
The CKM experiment requires $ 5 \times 10^{12}$ protons per second,
corresponding to less than 20 \% of the total design intensity of the 
Main Injector. 
Full scientific approval of CKM is expected in June 2001.

\subsection{$K_L \rightarrow \pi^0 \nu \bar{\nu}$}

This decay mode is considered to be the ``{\it holy grail of the kaon 
system}''~\cite{Littenberg}. Upper limits obtained so far lie 4 to 5
orders of 
magnitude above the SM prediction, as seen in
Table 2. 
A first generation of dedicated experiments is now being proposed. 
Comparisons of the experimental techniques  and of the 
overall expected sensitivities for currently 
proposed experiments are presented in Tables 3 and 4.
The two main difficulties of the experiment are the incomplete 
knowledge of the initial state and the daunting background 
originating from other $K_L$ and $\Lambda$ decays with $\pi^0$s and
neutron interactions.

\begin{center}
{\small Table 3: Experimental techniques for currently proposed  
$K_L \rightarrow \pi^0 \nu \bar{\nu}$ experiments.}
\vskip0.2cm
\begin{tabular}{|l|c|c|c c|}
\hline
& KEK-E391A & BNL-KOPIO & \multicolumn{2}{|c|}{FNAL-KAMI} \\
&           &           &   far       & near            \\
\hline
Calorimetry & CsI,CeF$_3$ & Shashlyk & \multicolumn{2}{|c|}{pure CsI} \\
$\sigma(E)/E (\%)$ & $0.4/\sqrt{E} + 0.5$ & $3.5/\sqrt{E}$ & 
\multicolumn{2}{|c|}{$2/\sqrt{E}\oplus 0.5$} \\
$\sigma(t)$ (ps) & & $50/\sqrt{E}$ & &  \\
$\sigma(\theta)$ (photons) & & 15 mrad& &  \\ 
\hline
Constraint from pencil beam ($p_{t}^{\rm max}$) (MeV/{\it c }) & 6 & &
6&11\\
\hline
Technique to measure $E_{K_L}$ & No & TOF& \multicolumn{2}{|c|}{No} \\
\hline
\begin{minipage}{55mm}{Required inefficiency \\ to veto photons 
( $E_{\gamma}~>~100$ MeV)}\end{minipage} & $<10^{-4}$ & $10^{-4}$ &
\multicolumn{2}
{|c|}{$<10^{-5}$}\\
\hline
kaon decay rate (MHz) & 0.5 & 25 & 2.8 & 8.2 \\
\hline
\end{tabular}
\end{center}

\begin{center}
{\small Table 4: Sensitivity for currently proposed 
$K_L \rightarrow \pi^0 \nu\bar{\nu}$ experiments.} 
\vskip0.2cm
\begin{tabular}{|l|c|c|c c|}
\hline
& KEK-E391A & BNL-KOPIO & \multicolumn{2}{|c|}{FNAL-KAMI} \\
&           &           &   far       & near            \\
\hline
Primary proton momentum (GeV/{\it c}) & 13 & 24 &
\multicolumn{2}{|c|}{120} \\
\hline
Production angle & 6$^\circ$ & 45$^\circ$ & \multicolumn{2}{|c|}{24
mrad}\\
\hline
Solid angle ($\mu$srd) & 16 & 500 & 0.36 & 1 \\
\hline
Protons on target (3y, 10$^7$ s/y) & $3.6\times  10^{19}$ & 
$6.1\times 10^{20}$ & \multicolumn{2}{|c|}{$3.3\times10^{20}$} \\
\hline 
Number of $K_L$ decays & $1.2\times 10^{12}$ & $1.5\times10^{14}$ &
$1.4\times10^{13}$ & $5.6\times10^{13}$ \\
\hline
Average $K_L$ momentum (GeV/{\it c}) &  2 & 0.7 & 13 & 10 \\
\hline
Acceptance (\%) & 10.2 & 1.5 & 7.1 & 7.4 \\
\hline
Signal events (BR $\simeq~3\times10^{-11}$) & 3 & 65 & 30 & 124 \\
\hline 
Background events & $<$ 1.8 & 35 & 17 & 40 \\ 
\hline
Foreseen data taking & 2001-2004 & 2004-2008& 2003-2005& 2006-? \\
\hline  
\end{tabular}
\end{center}

A significant improvement on the background rejection could be achieved
employing a beam of $K_L$ of known energy and direction, as could in
principle be provided by one of the following two
examples.
\begin{itemize}
\item{ A  $\phi$ factory:}

At a $\phi$ factory, the $K_L$ is produced at fixed momentum 
($\beta\simeq 0.21$) and its direction can be well determined 
using the reconstructed vertex for the accompanying 
$K_S \rightarrow \pi^+ \pi^-$  decay and the interaction
point~\cite{gino}. 
Another advantage in the search for $K_L \rightarrow \pi^0 \nu \bar{\nu}$
at a $\phi$ factory is the absence of background generated by 
$\Lambda \rightarrow n \pi^0$  decays and neutrons. 
\item{Exclusive formation by $\pi^- p \rightarrow K_L \Lambda$:}

This technique for producing $K_L$ of definite energy employing a 
1 GeV/{\it c} $\pi^-$ beam via the reaction $\pi^- p \rightarrow K_L
\Lambda$
has been used in the past~\cite{peach}.
\end{itemize}
Unfortunately, both methods appear to be orders of magnitude 
away from providing the necessary flux to reach SM sensitivities under 
reasonable running conditions.

The KEK-E391A~\cite{E391} and the KAMI ~\cite{KAMI} proposals exploit a 
pencil beam design 
to limit the kaon transverse momentum and apply cuts on the 
minimum $\pi^0$ transverse momentum. A cut on $p_{\rm T}~>~150$ MeV/{\it
c} 
effectively rejects backgrounds coming from the  
reaction $\Lambda \rightarrow \pi^0 n$.
To cope with the $K_L \rightarrow 2 \pi^0$ background with two lost
photons, 
the gamma veto inefficiency must be very small, $< 10^{-4}$ per photon.

A different approach is proposed by the KOPIO~\cite{KOPIO} experiment.
Here, the intention 
is to use a very wide beam extracted at large angle with respect to 
the primary proton beam. The 24 GeV/{\it c} primary proton beam can be 
delivered 
with 200 ps wide pulses spaced about 40 ns each~\cite{glenn}.
This microstructure
of the beam, coupled with the low kaon average momentum of 700 MeV/{\it
c},
allows measurement of the kaon momentum via Time Of Flight (TOF). The kaon 
decay can therefore be transformed into the centre-of-mass frame, 
providing a
constraint on the kinematics of the decay. 
In addition, the experiment plans to measure the photon direction by 
means of a preradiator placed in front of the electromagnetic
calorimeter. 

The performance of the calorimetry is stretched to the limits. Factors
such as
tails in the detector response should be considered in the evaluation 
of the response. As seen in Table 2,
there are 4 to 5 orders of magnitude to bridge from the current upper
limits to the SM prediction, and progress may be slower than expected. 

The MSR front end appears suitable for a second-generation experiment.  If
the current proposals do not reach the SM sensitivity, a second-generation
experiment should focus on improving the experimental technique. However,
one should plan an experiment capable of collecting the 1000 events which
would match the precision of the theoretical prediction. 

\subsection{$K_L \rightarrow \pi^0 e^+ e^-$}
The experimental difficulty of this decay is twofold. On the one side one
is forced to 
disentangle the short-distance direct CP violation from the indirect
CP-violating and CP-conserving ones. On the other hand, a large 
background from $K_L \rightarrow \gamma \gamma e^+ e^-$ cannot be reduced
below $10^{-10}$ without a significant efficiency loss. 
The best current limits are those presented by KTEV at Fermilab. 
Two candidtaes for the 
$K_L \rightarrow \pi^0 e^+ e^-$ decay remain in the signal box, 
for an expected background of $1.06\pm0.41$. The signal box is defined by
a $\pm 2.65$ MeV/c$^2$ ($\pm 5 {\rm MeV/c^2}$) cut on the
$m_{\gamma \gamma}$ ($m_{e^+ e^- \gamma \gamma}$) variables. These 
cuts correspond to a $\pm 2 \sigma$ cut on the signal distribution 
simulated by Monte Carlo.   
The limiting background is the radiative Dalitz decay $K_L \rightarrow 
e^+ e^- \gamma \gamma$ when $m_{\gamma \gamma} = m_{\pi^0}$. 
The most powerful variables for separating the signal from the background 
are the angle between the photons and the kaon in the $\pi^0$ rest frame 
and the minimum angle between any photon and any electron  in the kaon 
rest frame. 
The result presented by KTEV ~\cite{KTeV_p0ee} is based on data collected 
during the 1997 -- 1999 Fermilab fixed target run and corresponds to an 
exposure of $2.6 \times 10^{11}$ $K_L$ decays.  
The background is inversely proportional to the $m_{\gamma \gamma}$ 
resolution, which is dominated by the energy resolution of the
electromagnetic calorimeter.
The energy resolution for photons achieved by the KTEV CsI detector 
(cf. Table 3) represents the state of the art,
and a significant improvement over this figure is unlikely in the near 
future. Therefore the only hope to improve further on 
$K_L\to \pi^0 e^+ e^-$ is to measure an excess signal over the background. 
Progress will therefore only be made according to the square root of the 
available statistics. To address the SM prediction for 
direct CP violation in $K_L \rightarrow \pi^0 e^+ e^-$ about $10^{15}$
$K_L$ decays are needed. This corresponds to a flux which is about 20 
times larger than the exposure accumulated by the AGS-E871 experiment,
currently employing the most intense $K_L$ beam.

The $K_L \rightarrow \pi^0 \mu^+ \mu^-$ case is similar to the electron
channel: a smaller experimental background is offset 
by smaller values for the predicted branching fraction. 
In the short term, progress is expected from analysis of the data
collected
by KTEV in 1999. In the future, the KAMI experiment at Fermilab should 
be able to improve the present upper limits significantly.

\subsection{Lepton-flavour violation in kaon decays} 

The best limit to date on $K_L \rightarrow \mu e$ is that from the
AGS-E871 
experiment ~\cite{BNL871}, namely $B < 4.7 \times
10^{-12}$ at 90 \% 
CL. The neutral kaon beam was produced by 24 GeV/{\it c} protons on 
a 1.4 interaction length $Pt$ target. It was selected at an angle 
 3.75$^\circ$  with respect to the direction of the incoming protons. 
The typical proton intensity was 1.5$\times 10^{13}$ per spill 
of 1.2 - 1.6 s duration. About 2 $\times 10^8 K_L$ per spill with
momentum 
between  2 and 16 GeV/{\it c} entered the decay volume, 7.5 \% of which 
decayed in the 11 m fiducial volume. 
The undecayed neutral beam was absorbed by a beam stopper placed 
downstream of the first two tracking stations. 
The tracking detectors consisted of six chamber stations and two
consecutive 
dipole magnets, which had opposite polarities and provided
net transverse momenta of 418 and 216 MeV/{\it c}. 
The chamber stations where the highest rates occurred, close to 1 MHz rate
per wire, were 5 mm diameter straw tubes operated with a fast gas 
mixture~\cite{straw}. Electron identification was provided by an
atmospheric 
threshold Cherenkov counter and by a lead glass calorimeter. Muons 
were identified by scintillation counters and a range finder. 
The primary source of background was $K_L \rightarrow \pi e \nu$ decay in
which 
a pion decayed upstream of the muon filter. The misidentification of the
pion as a muon resulted in a maximum $m_{\mu e}$ of 489.3 MeV/$c^2$. The
$m_{\mu e}$ 
resolution was 1.38 MeV/$c^2$ and therefore background from Gaussian tails
is negligible. However, backgrounds from non-Gaussian tails could be
important. 
There is probably room for improvement by up to a factor of ten over this
result, but backgrounds from accidentals may become important for rates
ten times larger than E871.

\section{SOME CONSIDERATIONS ON USING THE PROTON DRIVER OF A\\
MUON STORAGE RING (MSR) AS A KAON FACTORY}

{\it G. Kalmus}

\subsection{Introduction}

The purpose of this section is to explore the feasibility of using the 
proton driver of a possible MSR facility to produce kaon beams of intensity 
and characteristics that are not only competitive with those available
elsewhere, but potentially even better.

\subsection{Assumptions}

\begin{figure}[t]
\hspace*{2.5cm}\epsfig{figure=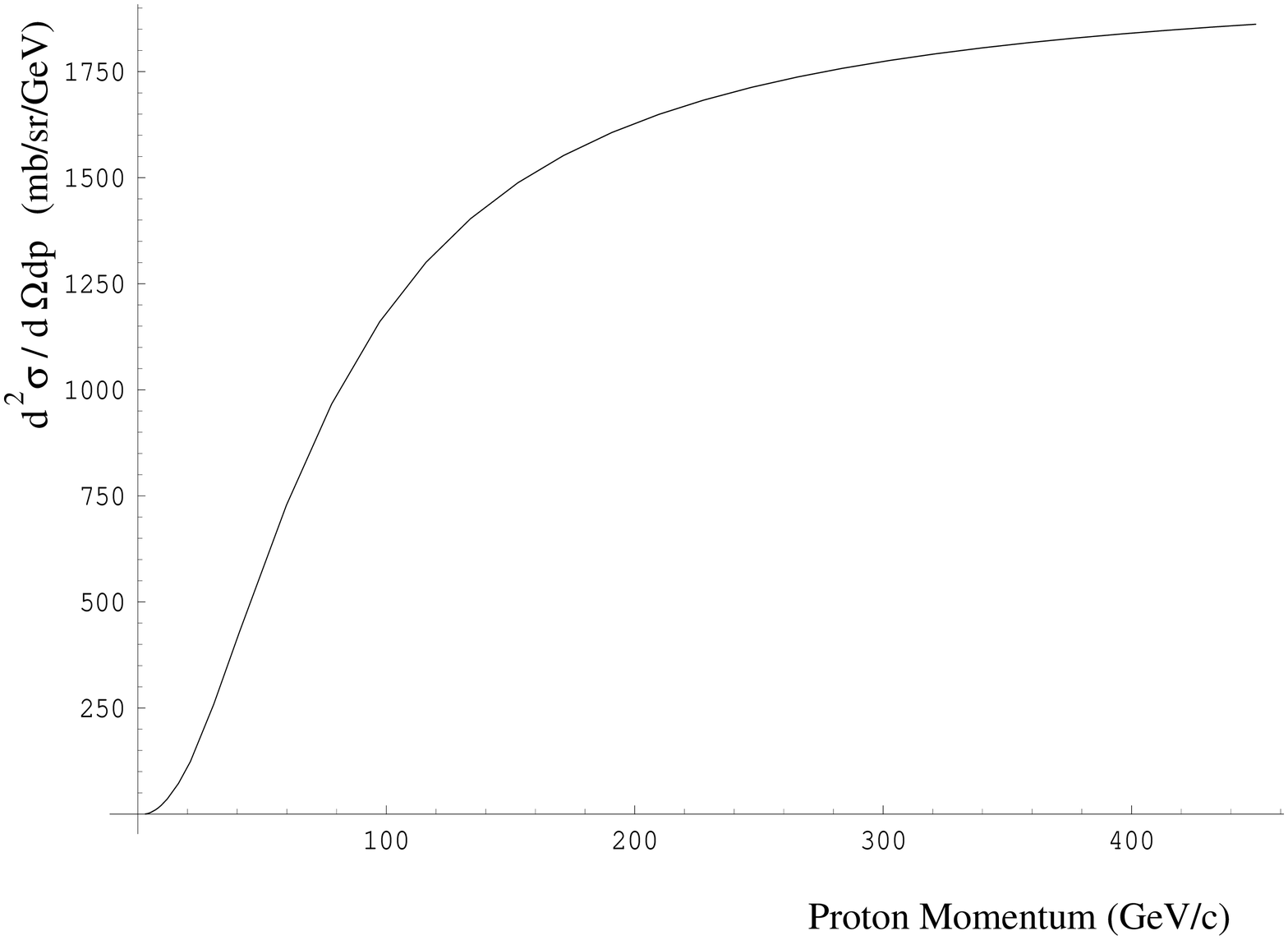,width=10cm,height=7cm}
\vspace*{0cm}
\caption{Differential $K_L$ production cross section as a function 
of the incident proton momentum. The cross section is for
a $Be$ target, in the forward direction ($\theta=0$), and with kaon
momenta integrated from 2 to 20 GeV. The estimate is based on the
empirical Sanford-Wang formula as described in \cite{AY}
(using $d\sigma(K_L)=(d\sigma(K^+)+3\, d\sigma(K^-))/4$).}
\label{fig:klsig220}
\end{figure}

\begin{figure}[h!]
\vspace*{-2cm}
\hspace*{2.5cm}\epsfig{figure=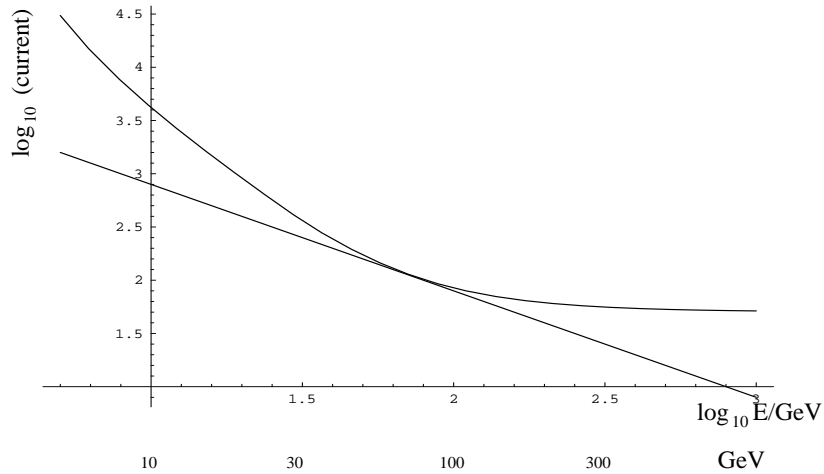,width=11cm,height=12cm}
\vspace*{-2cm}
\caption{The beam current, in arbitrary units, is plotted against
the beam energy in GeV. The straight line represents a curve of
constant beam power. The convex line is a curve of constant kaon yield,
based on the same input as assumed for Fig. \ref{fig:klsig220}.
Since the accuracy of the empirical formulae is limited, this plot
should mainly be understood as a qualitative illustration. However,
this plot indicates that the beam power required for a given kaon
yield would be lowest at energies of 30 to 100 GeV.}
\label{fig:klyield}
\end{figure}

The qualitative behaviour of kaon production as a function
of machine parameters (beam current and energy) is shown
in Figs.~\ref{fig:klsig220} and \ref{fig:klyield}.
We see in Fig.~\ref{fig:klsig220} that the kaon yield rises rapidly as a
function of energy: the $K_L$ production cross section shown is
expected to be similar to the mean of the $K^+$ and $K^-$ yields.
On the other hand, the absolute yield is not necessarily the most
important
consideration. One might also wish to optimize the yield {\it for a
given beam power}, which is shown in Fig.~\ref{fig:klyield}. However,
economic
and other considerations might motivate a choice of proton driver energy
somewhat below the optimal range between 30 and 100 GeV. For the purposes
of this brief study, we make the following default assumptions.

\noindent
a) We assume that the MSR proton driver achieves 24 GeV, and is a
rapid-cycling synchrotron with a 3$\mu$s beam pulse operating at 15 Hz.

\noindent
b) The total beam power is 4MW, giving a beam current of about 160$\mu$A.

\noindent
c) A stretcher ring is available and able to convert the very bad 
duty cycle ($< 10^{-4}$) to a good one ($\sim 100\%$).

\noindent
d) Slow ejection from the stretcher ring at these high intensities is 
possible and efficient.

\noindent
e) $10\%$ of the protons from the MSR source can be used for kaon physics,
i.e., 3 pulses every 2 seconds can be routed to the stretcher rings.

It is clear from Figs.~\ref{fig:klsig220} and \ref{fig:klyield} that a
2.2~GeV proton driver, being close to the kaon production threshold, would
be vastly inferior as a kaon source, and not competitive. However, even in
this case, one could still consider the possibility of post-accelerating
$10\%$ of these low-energy protons to 24 GeV.

\subsection{Machine parameters and secondary beams}

Table 5 gives the parameters for proton machines that are existing,
are under construction, or have been proposed.
It should be noted that these are numbers for a full-intensity beam. In
the case of the MSR proton driver, for the reasonable assumption of 3
pulses every
2 seconds, i.e., 10\% of the beam, the cycle rate and protons/sec should
be divided by 10. 
\begin{center}
{\small Table 5: Comparison of proton intensities from existing, 
projected and proposed machines.}
\vskip0.2cm
\begin{tabular}{|c|c|c|c|c|c|}
\hline
& Beam Energy & Beam Current & Cycle Rate & p/Pulse & p/sec \\
& (GeV)       &   ($\mu$A)   & Hz         &         &       \\
\hline
CERN PS & 26 & 1.6 & 0.5 & $2\times 10^{13}$ &  $10^{13}$ \\
\hline
BNL AGS & 24(30) & $\sim 5$ & 0.3 & $10^{14}$ & $3\times 10^{13}$ \\
\hline
FNAL MI (2002?) & 120 & 1.6 & 0.33 & $3\times 10^{13}$ & $10^{13}$ \\
\hline
JHF (2006/7?) & 50 & 10 & 0.16 & $4\times 10^{14}$ & $6\times 10^{13}$ \\
\hline
KEK & 12 & 0.16 & 0.25 & $4\times 10^{12}$ & $10^{12}$ \\
\hline
KAON (Defunct) & 30 & 100 & 10 & $6\times 10^{13}$ & $6\times 10^{14}$ \\
\hline
CERN MSR source & 24 & 160 & 15 & $7\times 10^{13}$ & $10^{15}$ \\
(201N?) & & & & & \\
\hline
\end{tabular}
\end{center}

Table 6 gives the characteristics of beams available at the Brookhaven AGS 
operating at 24 GeV~\cite{DLP}.
\begin{center}
{\small Table 6: Kaon beams available from the Brokkhaven AGS.}
\vskip0.2cm
\begin{tabular}{|c|c|c|c|c|c|c|c|c|}
\hline
Beam & Kaon & $\delta p/p$ & Prod. & $\Delta\Omega$ 
           & \multicolumn{2}{l|}{Flux} & Purity & Rem. \\
& Mom. & FWHM & Angle &  
           & \multicolumn{2}{l|}{per $10^{13}$ p} & & \\
& (GeV/c) & (\%) & (deg) & (msr) & \multicolumn{2}{c|}{} & & \\
\hline
& & & & & $K^+$ & $K^-$ & & \\
\hline
C4 & $\stackrel{<}{_\sim}0.83$ & 4 & 0 & 12 & 4.6 & 1.5 
   & $\frac{\pi^+}{K^+}=0.4$ & L=18m \\
(separated) & & & & & $\times 10^6$ & $\times 10^6$ & & 
             $\sim 10^6 K^+$ \\
& & & & & & & & stopped \\
& & & & & & & & per $10^{13}$ p \\
\hline
D6 & $\stackrel{<}{_\sim}1.9$ & 6 & 5 & 1.6 & 5.5 & 2.3 
   & $\frac{\pi^-}{K^-}=0.8$ & L=31m \\
(separated) & & & & & $\times 10^6$ & $\times 10^6$ & & \\
\hline
B5 & 2--20 & & 1--4.5 & 0.1 & \multicolumn{2}{l|}{$K^0_L$ flux @ $3.75^\circ$} 
           & $\frac{n}{K^0_L}=20$ & L=10m \\
(neutral) & & & & & 
          \multicolumn{2}{l|}{$\sim 1.3\times 10^8$} & & \\

\hline
\end{tabular}
\end{center}

Since the MSR proton driver that we assume has a similar energy to that of
the BNL AGS, it is easy to estimate the intensities of analogous beams
produced by the proton driver for the MSR, using 1/10 of its intensity for
a Kaon beam.

\noindent
a) The C4 beam would deliver:           
\begin{eqnarray}
\left.
\begin{array}{c}
4.6\times 10^7 K^+/{\rm sec}\\
1.5\times 10^7 K^-/{\rm sec}
\end{array}\right\}  && {\rm at~0.8~GeV/c}\\
1\times 10^7 K^+/{\rm sec} &&  {\rm at~rest}
\end{eqnarray}

\noindent
b) The D6 beam would deliver:   
\begin{eqnarray}
\left.
\begin{array}{c}
5.5\times 10^7 K^+/{\rm sec}\\
2.3\times 10^7 K^-/{\rm sec}
\end{array}\right\}  && {\rm at~1.8~GeV/c}
\end{eqnarray}

\noindent
c)  The B5 beam would deliver:  
\begin{equation}
1.3\times 10^9 K_L/{\rm sec} {\rm ~in~the~range~2-20~GeV}
        {\rm ~with~}  n/K^0_L\sim 20
\end{equation}

\subsection{Summary}

On the basis of this brief survey, we consider the following points to be 
established.

\noindent
a) A high-energy ($> 10$~GeV) MSR source is needed to produce competitive 
kaon beams.

\noindent
b) For the MSR source assumed here, with an energy of 24 GeV and a beam
power 
of 4MW, a $10\%$ share of the protons will produce about 10 times the 
flux of kaons produced by $100\%$ of the CERN PS or 3 times the flux 
produced by $100\%$ of the Brookhaven AGS.

\noindent
c) In order to be able to utilize these high fluxes, even in separated 
beams, the short, rapid cycling synchrotron pulses must be stretched
so that the duty cycle is increased from $\sim 10^{-4}$ to $\sim 1$.

\noindent
d) For neutral beams with $10^9 K_L/$sec and 
$2\times 10^{10}$ neutrons/sec, very high 
rates will be recorded in detectors.

\noindent
e) The beam extraction, targeting and experiments will need to be very
carefully 
designed.

\noindent
f) If these technical problems can be overcome, then $10\%$ of a 24-GeV 
MSR source could provide intense kaon beams at low energies, e.g.,
$\sim 5\times 10^{13}$ stopping $K^+$/year or
$\sim 5\times 10^{14}$ $K^0_L$/year decaying over a 15m length.


\section{SUMMARY AND CONCLUSIONS}

{\it G. Buchalla, A. Ceccucci}

We have studied opportunities for precision experiments in
kaon physics using a high-intensity proton driver at a future
muon storage ring (MSR) facility.
Rare decays of kaons are excellent tools to test the flavour sector
in great detail, to determine precisely the CKM matrix and
to search with high sensitivity for signatures of new physics.
We have stressed a number of highlights that combine outstanding
physics motivation with the need of a longer term experimental
effort towards the necessary measurements. As such they are
suitable targets for a possible MSR front-end facility.

\begin{itemize}

\item
$K_L\to\pi^0\nu\bar\nu$: This rare decay mode is the benchmark
process for determining the Jarlskog parameter $J_{CP}\sim\eta$,
the invariant measure of CP violation in the standard model (SM).
On the order of 1000 clean events could be used before being limited
by the very small theoretical uncertainties. 
Simultaneously,
$K_L\to\pi^0\nu\bar\nu$ is an excellent new physics probe.

\item
$K^+\to\pi^+\nu\bar\nu$: This related CP-conserving process is
sensitive to $|V_{td}|$ and has only slightly larger 
theoretical uncertainties than the neutral mode. It also
represents a very important goal for future kaon experiments.

\item
$K_L\to\pi^0e^+e^-$: This process has a substantial
contribution from direct CP violation, which is analogous to the
$K_L\to\pi^0\nu\bar\nu$ amplitude and likewise has very small
theoretical uncertainties. Moreover, the sensitivity to new physics
is quite different for the two cases, so that a measurement of 
$K_L\to\pi^0e^+e^-$ is not redundant, but will provide additional
information. However, in contrast to the neutrino mode, virtual
photons can induce two more contributions in the case of
$K_L\to\pi^0e^+e^-$, one from indirect CP violation proportional
to $\varepsilon_K$ times the $K_S\to\pi^0e^+e^-$ amplitude,
and also a CP-conserving piece adding incoherently to the rate.
All three contributions have to be disentangled to extract the
most interesting part from direct CP violation.
For this reason the phenomenology of $K_L\to\pi^0e^+e^-$ is more
complicated than it is in the case of $K_L\to\pi^0\nu\bar\nu$.
On the other hand, the prospects of kaon beams of very high intensity
suggest the possibility of resolving these problems 
using various approaches that make use of sufficiently high
statistics measurements. Building on previous proposals,
we have therefore performed a new phenomenological analysis of
$K_L\to\pi^0e^+e^-$, with a high-intensity kaon facility in mind and
taking into account recent developments and results.
New features in this study include in particular:

\vspace*{0.2cm}

\begin{itemize}
\item
The prospects of experimental results on $B(K_S\to\pi^0e^+e^-)$
in view of a heuristic theoretical estimate that this branching
fraction may be large. This experimental input is crucial for analysing
$K_L\to\pi^0e^+e^-$.
\item
Methods to extract direct CP violation depending on the size of
$B(K_S\to\pi^0e^+e^-)$, especially if it is large.
\item
New experimental results on  $K_L\to\pi^0\gamma\gamma$ and a 
re-analysis of the CP-conserving contribution, which appears
to be well under control.
Suitable cuts can further reduce the CP-conserving component
in an efficient way.
\item
Quantitative analysis of time-dependent $K_L$--$K_S$ interference
in $K\to\pi^0e^+e^-$ decays, including the impact of a
CP conserving contribution and the Greenlee background
($K_L\to e^+e^-\gamma\gamma$).
\end{itemize}

\vspace*{0.2cm}

For the proposed measurements in $K\to\pi^0e^+e^-$ to be
interesting, the typical number of required decaying $K_L$ is
$\sim 10^{15}$ per year.

\item
Lepton-flavour violation in $K_L\to\mu e$,
$K\to\pi\mu e$:
These processes are of special interest as they are absent in the
SM, and thus serve as direct indicators of new physics.
In general, these decays are often constrained in a given model
both by bounds on $\mu\to e$ transitions as well as by flavour
violation in the quark sector, particularly $\Delta m_K$. However, since
decays such as $K_L\to\mu e$ involve flavour violation simultaneously
in the quark and lepton sectors, direct measurements of these decays 
can in general give complementary information.
An important scenario is provided by supersymmetric theories. In the MSSM
the constraints from $\mu\to e$ transitions and $\Delta m_K$ are
generally very tight, but $K_L\to\mu e$ branching ratios accessible
to future high-intensity kaon experiments could still be allowed for
certain regions of parameter space. Much larger effects just below 
current bounds could await discovery in the more general case of 
supersymmetry with $R$-parity violation.
Additional interesting probes of LFV are decays of the type 
$K^+\to\pi^-\mu^+\mu^+$.
\end{itemize}

We have further reviewed the status and prospects of current and planned
rare $K$ decay experiments that are relevant to the processes of
special interest for this study. Possible improvements and strategies
towards reaching the necessary higher sensitivity have been suggested.

We have also outlined machine requirements needed to realize the
potential of kaon physics in the context of a MSR front end.
The high-intensity proton source should have an energy of at least
$10$~GeV. As a benchmark, we have considered a
proton machine with 24 GeV energy and 160 $\mu$A beam current.
A useful scenario for the purpose of comparison is given
by the proposal of a European Hadron Facility (EHF) described
in~\cite{EHF}.
 
\vspace{0.7cm}

In conclusion, rare decays of kaons, such as
$K_L\to\pi^0\nu\bar\nu$, $K_L\to\pi^0e^+e^-$ and $K_L\to\mu e$,
present excellent opportunities to obtain competitive and complementary
insight into flavour physics, one of the forefront areas in high-energy
research. Current or planned near-future experiments will not fully
exploit the physics potential of these processes.
High-intensity kaon beamlines made available by a muon storage ring
complex could allow us to perform the necessary
second-generation kaon experiments.

\vskip1cm
\noindent

\subsection*{Acknowledgements}

We thank Bruno Autin, Giles Barr, Alain Blondel, Lau Gatignon,
Helmut Haseroth, Takao Inagaki, Miko\l{}aj Misiak, Ulrich Nierste,
Francesco Pietropaolo, Guennadi Volkov and Akira Yamamoto
for useful discussions and contributions to the working group.

\end{document}